\documentclass[amssymb,amsmath,superscriptaddress,footinbib]{revtex4}
\usepackage{natbib,graphicx,subfigure,subeqnarray,fancyhdr,amsmath,multirow,color, mathrsfs}
\begin{document}

\renewcommand{\labelitemi}{$-$}
\newcommand{\change}[1]{{\color{red} #1}}
\newcommand{\Fc}{\mathcal{F}}\newcommand{\Rc}{\mathcal{R}}\newcommand{\dd}{\mathrm{d}}
\newcommand{\ee}{\mathrm{e}}\newcommand{\ci}{\mathrm{i}}\newcommand{\ib}{\mathbf{i}}
\newcommand{\jb}{\mathbf{j}}\newcommand{\kb}{\mathbf{k}}\newcommand{\ab}{\mathbf{a}}
\newcommand{\Fb}{\mathbf{F}}\newcommand{\fb}{\mathbf{f}}\newcommand{\Gb}{\mathbf{G}}
\newcommand{\Mb}{\mathbf{M}Ä}\newcommand{\nb}{\mathbf{n}}\newcommand{\Sb}{\mathbf{S}}
\newcommand{\Sbs}{\mathbf{S^*}}\newcommand{\Rb}{\mathbf{R}}\newcommand{\Sigb}{\boldsymbol{\Sigma}}
\newcommand{\Sigbs}{\boldsymbol{\Sigma^*}}
\newcommand{\omegab}{\boldsymbol{\omega}}
\newcommand{\epsb}{\boldsymbol{\epsilon}}
\newcommand{\ub}{\mathbf{u}}
\newcommand{\eb}{\mathbf{e}}\newcommand{\vv}[1]{\underline{#1}}\newcommand{\ev}{\vv{e}}
\newcommand{\rv}{\vv{r}}\newcommand{\TT}[1]{\underline{\underline{#1}}}\newcommand{\omb}{\mathbf{\omega}}
\newcommand{\Ub}{\mathbf{U}}\newcommand{\xb}{\mathbf{x}}\newcommand{\rb}{\mathbf{r}}
\newcommand{\ssb}{\mathbf{s}}\newcommand{\Xb}{\mathbf{X}}\newcommand{\Pe}{\mbox{Pe}}
\newcommand{\mean}[1]{\langle #1\rangle}
\newcommand{\ddp}{[p]^\pm}\newcommand{\taub}{\mbox{\boldmath$\tau$}}\newcommand{\Fr}{\mbox{\textit{Fr}}}
\let\grad\nabla\newcommand{\z}{\zeta}\newcommand{\kk}{\kappa}\newcommand{\tkk}{\tilde{\kappa}}
\newcommand{\e}{\varepsilon}\newcommand{\zb}{\bar{\zeta}}\let\grad\nabla\let\bcdot\cdot
\newcommand{\half}{{\textstyle\frac{1}{2}}}
\newcommand{\textfrac}[2]{{\textstyle\frac{#1}{#2}}}
\newcommand{\LF}[1]{{#1}^{\mathrm{LF}}}\newcommand{\Lap}[1]{{#1}^{\mathrm{L}}}
\newcommand{\ds}{*\!*}\newcommand{\cond}[2]{\frac{\mathrm{D} #1}{\mathrm{D} #2}}
\newcommand{\pard}[2]{\frac{\partial #1}{\partial #2}}\newcommand{\totd}[2]{\frac{\mathrm{d}#1}{\mathrm{d}#2}}
\newcommand{\pardd}[3]{\frac{\partial^2 #1}{\partial #2 \partial #3}}
\newcommand{\Rey}{\mbox{Re}}\newcommand{\Imag}{\mbox{Im}}
\newcommand{\Fpint}{=\!\!\!\!\!\!\!\int}
\newcommand{\txi}{\tilde\xi}\newcommand{\dxi}{\delta\xi}
\newcommand{\tpsi}{\tilde\psi}\newcommand{\dpsi}{\delta\psi}
\makeatletter
\def\sgn{\mathop{\operator@font sgn}}
\makeatother

\title{Unsteady feeding  and optimal strokes of model ciliates}
\author{S\'ebastien Michelin}
\email{sebastien.michelin@ladhyx.polytechnique.fr}
\affiliation{LadHyX -- D\'epartement de M\'ecanique, Ecole
  polytechnique, 91128 Palaiseau Cedex, France.}
\author{Eric Lauga}
\email{elauga@ucsd.edu}
\affiliation{Department of Mechanical and Aerospace Engineering, University of California San Diego, 9500 Gilman Drive, La Jolla CA 92093-0411, USA.}

\date{\today}

\begin{abstract}

The flow field created by swimming microorganisms not only enables their locomotion but also leads to advective transport of nutrients. In this paper we address analytically and computationally the link between unsteady feeding and unsteady swimming on a model microorganism, the spherical squirmer, actuating the fluid in a time-periodic manner. We start by performing asymptotic calculations at low  P\'eclet number ($\Pe$) on the advection-diffusion problem for the nutrients. We show that the mean rate of feeding as well as	 its  fluctuations in time depend only on the swimming modes of the squirmer up to order $\Pe^{3/2}$, even when no swimming occurs on average, while the influence of  non-swimming modes comes in  only at order $\Pe^2$. We also show that generically we expect a phase delay between feeding and swimming of 1/8th of a period. Numerical  computations for illustrative strokes  at finite $\Pe$ confirm quantitatively our analytical results  linking  swimming and feeding.  We finally derive, and use, an adjoint-based optimization algorithm  to   determine the optimal unsteady strokes maximizing feeding rate for a fixed energy budget. The overall optimal feeder is always the optimal steady swimmer. Within the set of time-periodic strokes, the optimal feeding strokes are found to be equivalent to those optimizing periodic swimming  for all values of the P\'eclet number, and correspond to a regularization of the overall steady optimal.
\end{abstract}

\maketitle

\section{Introduction}

In order to be able to swim in viscous fluids, micro-organisms must undergo non-time-reversible sequences of shape changes referred to as swimming strokes \citep{lighthill1975,purcell1977,lauga2009}. Through the no-slip boundary condition, these  strokes induce a net  flow field  around the organism and a distribution of viscous stresses which lead to locomotion. This swimming-induced flow  also impacts hydrodynamic interactions with neighboring organisms \citep{drescher2009,michelin2010a} or material boundaries \citep{berke2008,lauga2006}, the overall dynamics of suspensions of cells \citep{kessler1986,pedley1992,sokolov2007,saintillan2008a,evans2011} and the feeding ability of organisms \citep{childress1987,short2006}.

Cellular motility is essential to many biological functions, from reproduction \citep{suarez2006} to escaping agressions \citep{crawford1992,hamel2011}. It also allows organisms to travel toward better local environments for example to seek (or escape) light, nutrient, or heat. The performance of the particular stroke displayed by  a single micro-organism, or that of a suspension of such swimmers, also results in the modification of the bulk stress and effective viscosity of a flow \citep*{batchelor1970,ishikawa2007a}, or of its mixing properties \citep{saintillan2008b,leptos2009,kurtuldu2011}, an effect that is suspected to play an important role on large-scale bio-mixing in the ocean for example \citep*{doostmohammadi2012}.

The metabolism of many microorganisms relies on the absorption of diffusing nutrients present in their vicinity, ranging from dissolved gases and low-weight proteins, to more complex molecular compounds and, in the case of large organisms such as the protozoon \emph{Paramecium}, smaller bacteria whose run-and-tumble motion is equivalent to a diffusive process at the scale of \emph{Paramecium} \citep{berg1993}. For a particular microorganism, the impact of the stroke on its feeding ability can be thought of as  twofold: (i) through  the motility resulting from the stroke, the organism can travel toward nutrient-rich regions;  (ii) by stirring nutrients in its immediate vicinity, the stroke-induced flow modifies, and possibly enhances,  local concentration gradients.  

The competition of advective and diffusive effects on the dynamics of a particular nutrient is quantified in the P\'eclet number, $\Pe=\tau_\textrm{diff}/\tau_\textrm{adv}$, where $\tau_\textrm{diff}=a^2/\kappa$ and $\tau_\textrm{adv}=a/\mathcal{U}$ are the characteristic diffusive and advective time-scales respectively, where $a$, $\mathcal{U}$ and $\kappa$ are the typical size of the organism, the characteristic flow velocity, and the nutrient diffusivity, respectively. Depending on the nutrient considered, $\Pe$ can vary by several orders of magnitude, even for a given microorganism.

Performing its stroke represents an energetic cost for the organism, as it must work against the fluid to overcome viscous dissipation. How far it can swim or how much nutrient it can absorb is therefore, in theory, limited by the finite amount of energy it has available. Considering that energy losses other than hydrodynamic can be accounted for by a fixed metabolic efficiency, the optimization of the swimming stroke to maximize either motility or feeding  can therefore be formulated as follows: for a fixed amount of energy available to deform its shape, what is the optimal stroke of a particular micro-organism maximizing either (i) the net displacement (optimal swimming problem) or (ii) the amount of a particular nutrient absorbed at the surface of the organism (optimal feeding problem)? In the latter case, the optimal stroke does not necessarily require a net displacement of the cell, as the organism can potentially just sit in a given location and stir the fluid around it. The optimal feeding stroke may also depend on the particular nutrient considered and the relative importance of advection and diffusion through the value of $\Pe$.

The optimization problems described above are closely linked to the question of optimality with respect to a specific biological function, which can take two different forms: optimal shape or optimal gait. In the former, one is interested in the optimal morphology of the swimmer (e.g. its aspect ratio, the use of flagella vs. cilia,...) and compares different species of microorganisms. In the latter, the focus is placed on a given organism, and the goal is to determine the sequence of body deformations that performs best \citep{tam2007,spagnolie2010,michelin2010c,michelin2011,tam2011,tam2011b}.

In this work, we focus on the optimal gait of a particular swimmer model, the so-called squirmer. This canonical model, consisting of a spherical microorganism imposing a tangential velocity at its surface, was introduced as a so-called enveloppe model for ciliated microorganisms \citep{lighthill1952,blake1971}. Ciliates, such as \emph{Paramecium}, swim in viscous flows using the coordinated beating of a large number of small cilia distributed over their surface \citep{blake1974b,brennen1977}.  In the squirmer model, the flow field can be determined analytically through the projection of the stroke on orthogonal squirming modes. Because of its simplicity, this model has been used to study a large variety of problems related to swimming microorganisms, including hydrodynamic interactions \citep*{ishikawa2006}, mixing \citep*{lin2011}, suspension rheology \citep{ishikawa2007b}, collective dynamics and instabilities \citep{ishikawa2007a,evans2011}, and feeding \citep{magar2003,magar2005,doostmohammadi2012}. 

Recently, \citet{michelin2010c} determined the optimal time-periodic swimming strokes (i.e. those maximizing the swimming velocity for fixed energetic cost) of such a model microorganism, and identified their main properties. In a subsequent contribution, \citet{michelin2011} considered the optimization of the stroke for feeding in the particular case of a steady surface velocity. Although such strokes correspond to non-periodic displacements of the surface, the results shed some light on the link between swimming and feeding, and in particular it was shown that  optimal swimming strokes and optimal feeding strokes were essentially identical regardless of $\Pe$, a result that is not a priori intuitive due to the fundamental differences in the impact of swimming on feeding at low or high $\Pe$: at low $\Pe$, swimming only impacts marginally the nutrient distribution, but enables the organism to travel toward regions with richer nutrient content, while at high $\Pe$, swimming also impacts feeding through stirring and strong advection of the nutrient in the vicinity of the organism surface.

The validity of these conclusions, and in particular the intimate relationship between optimal swimming and optimal feeding, remains however to be addressed in the  general  case of unsteady strokes. \citet{magar2005} showed that in the particular limit of large $\Pe$ and small surface displacement, an equivalent steady problem could be defined. However, the unsteady effects of advection and diffusion in the general case of both finite swimmer displacement and finite $\Pe$ number remain unclear. In this paper, we specifically focus on the unsteady swimming problem. We first  address analytically and computationally the link between unsteady feeding and unsteady swimming.  We then derive, and use, an adjoint-based optimization algorithm  to   determine the optimal unsteady strokes maximizing feeding rate for a fixed energy budget.

The paper is organized as follows. In \S\ref{sec:model}, the squirmer model is briefly presented, and the swimming and feeding problems are posed mathematically. In \S\ref{sec:asymptotics}, the unsteady feeding rate is determined in the asymptotic limit of small $\Pe$. The impact of the swimming stroke and of the P\'eclet number on the feeding rate is further analyzed in \S\ref{sec:unsteady_simulations} using numerical simulations, providing an important insight on the link between swimming and feeding. Section~\ref{sec:optimization} presents the result of the stroke optimization with respect to feeding and conclusions and perspectives are finally presented in \S\ref{sec:conclusions}.

\section{Swimming and feeding of a model ciliate}
\label{sec:model}
\subsection{The squirmer model}
The present work focuses on a particular model micro-organism, the squirmer, illustrated in  figure~\ref{fig:squirmer}. It is a spherical organism of radius $a$ which prescribes  periodic tangential deformations of its surface $\mathcal{S}$ with a frequency $\omega$, in order to swim in a viscous fluid of dynamic viscosity $\mu_f$ and density $\rho_f$. The present analysis is restricted to purely axisymmetric deformations of $\mathcal{S}$ so that the swimming velocity is parallel to the axis of symmetry $\eb_x$, with no rotation. In this paper, we will seek optimal strokes maximizing the  feeding rate of the organism for a given amount of energy available during each period to perform its surface deformation (and possibly its swimming). This average rate of energy consumption, $\mathscr{P}$, is identified with the rate of work applied on the fluid by the swimmer at its surface, or, equivalently, the total mechanical energy dissipated in the fluid through viscous effects during one period. It is related to the typical surface velocity scale $\mathscr{U}$ by
\begin{equation}
\mathscr{U}=\sqrt{\frac{\mathscr{P}}{12\pi\mu_f a}}\cdot
\end{equation}

The squirmer is swimming in a continuous suspension of a given nutrient (e.g. bacteria, large proteins/molecules, heat...) characterized by a far-field concentration $C_\infty$ and a diffusivity $\kappa$, and advected by the flow created by the surface stroke. On the swimmer boundary, the nutrient is instantaneously absorbed and processed at the surface so that $C=C_b$, with $C_b$ the equilibrium concentration at the surface determined by the processing mechanism. A more realistic, but more complex, boundary condition was proposed by \citet{magar2003} and \citet{magar2005}, taking into account such effects as the resistance of the membrane to nutrient absorption, and the finite diffusion and processing time of the nutrient within the cell.  The instantaneous nutrient uptake by the organism through diffusion at its boundary, $\Phi(t)$,  is given by 
\begin{equation}
\Phi(t)=\int_\mathcal{S}\kappa\pard{C}{r}\dd S.
\end{equation}
In the case of a purely rigid sphere, with no advection, a steady nutrient flux is achieved through diffusion $\Phi_0=4\pi a\kappa (C_\infty-C_b)$. In the following, we focus on the modification of the concentration field by the organism and define the rescaled concentration field $c=(C_\infty-C)/(C_\infty-C_b)$. 

Three distinct time-scales are present in the problem: (i) a diffusive time-scale $\tau_d=a^2/\kappa$, (ii) an advective time-scale $\tau_a=a/\mathscr{U}$ and (iii) the stroke period $\tau_\omega=2\pi/\omega$, while only the latter two were present in the purely swimming problem \citep{michelin2010c} and only the first two in the steady feeding problem \citep{michelin2011}. 
 The P\'eclet number, $\Pe=\tau_d/\tau_a$, is a measure of the relative importance of advective and diffusive effects near the surface of the squirmer, and is equal to
\begin{equation}
\Pe=\frac{\mathscr{U}a}{\kappa}=\frac{1}{\kappa}\sqrt{\frac{\mathscr{P}\,a}{12\pi\mu_f}}\cdot
\end{equation}
A second independent time-scale ratio can be defined either as a characteristic of the stroke, for example the relative velocity $U_R=\mathscr{U}/(a\omega)$, or as a period-based P\'eclet number $\Pe_\omega=a^2\omega/\kappa$. 
In the following, all equations and quantities are non-dimensionalized using $a$, $\omega$, $\mu_f$, and $C_\infty-C_b$ as reference quantities.

\subsubsection{Swimming problem}
\begin{figure}
\begin{center}
\includegraphics[width=.65\textwidth]{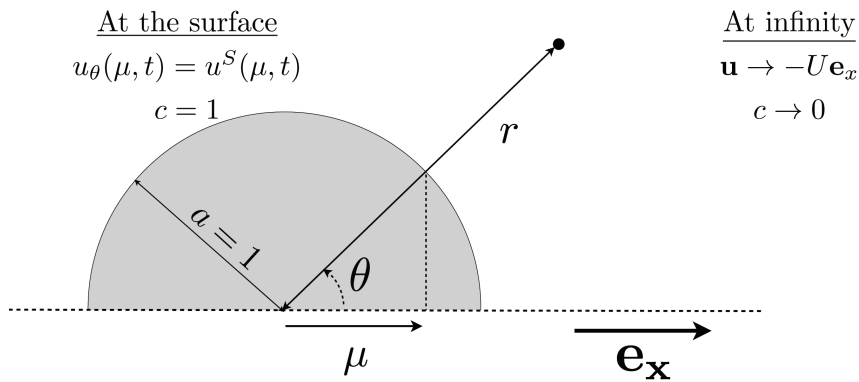}
\caption{Swimming and feeding of a squirmer. A purely axisymmetric tangential velocity and a purely absorbing boundary condition for the nutrient are imposed at the surface of the swimmer. All variables are non-dimensional.}\label{fig:squirmer}
\end{center}
\end{figure}
 Due to the small size of the organisms considered, the Reynolds number, $\Rey=\rho_f\mathscr{U} a/\mu_f$, a relative measure of inertia and viscous effects in the flow, is always much smaller than one, and the velocity and pressure fields satisfy Stokes equations. The swimming problem in the reference frame attached to the organism is therefore
 \begin{equation}\label{eq:stokes}
 \nabla^2\ub=\grad p,\quad \grad\cdot\ub=0,
 \end{equation}
 with the boundary conditions on the swimmer surface and at infinity given by
 \begin{eqnarray}
& \ub=u_\theta^S(\mu,t)\eb_\theta&\quad \textrm{at  } r=1,\\
& \ub\rightarrow -U(t)\eb_x&\quad \textrm{for  } r\rightarrow\infty.
 \end{eqnarray}
Note that the prescribed surface   field, $u_\theta^S$, is the stroke imposed by the organism and at the origin of both locomotion and stirring.  The stroke is assumed to be axisymmetric, therefore the surface velocity only depends on $\mu=\cos\theta$ and $t$, with $\theta$ the polar angle measured from the swimming direction $\eb_x$ (figure~\ref{fig:squirmer}). In Stokes flow, the swimmer can not sustain any net hydrodynamic force, therefore we have 
 \begin{equation}\label{eq:zeroforce}
 \int_\mathcal{S}\left[-p\boldsymbol{1}+\left(\grad\ub+\grad\ub^T\right)\right]\cdot\nb\,\dd S=0,
 \end{equation}
 where $\nb$ is the unit normal vector pointing into the fluid ($\nb=\eb_r$ here). Note that we have assumed the swimmer to be neutrally buoyant.   The solution to the swimming problem in~\eqref{eq:stokes}--\eqref{eq:zeroforce} is obtained by decomposing the surface velocity onto the squirming modes \citep{blake1971,michelin2010c}
 \begin{equation}
 u_\theta^S(\mu,t)=\sum_{n=1}^\infty\alpha_n(t)K_n(\mu),
 \end{equation}
 with
 \begin{equation}
K_n(\mu)=\frac{2n+1}{n(n+1)}\sqrt{1-\mu^2}L_n'(\mu),
 \end{equation}
 where $L_n(\mu)$ is the $n$-th order Legendre polynomial. The values of the pressure field and streamfunction are then obtained as
 \begin{align}
 p(r,\mu,t)&=p_\infty+\sum_{n=2}^\infty\alpha_n(t)P_n(r,\mu),\\
 \psi(r,\mu,t)&=\sum_{n=1}^\infty\alpha_n(t)\Psi_n(r,\mu),\label{eq:stream}
 \end{align}
 with
 \begin{align}
 P_n(r,\mu)&=-\left(\frac{4n^2-1}{n+1}\right)\frac{L_n(\mu)}{r^{n+1}},\\
 \Psi_n(r,\mu)&=\frac{2n+1}{n(n+1)}(1-\mu^2)L_n'(\mu)\psi_n(r),\label{eq:streamn}\\
 \psi_1(r)=\frac{1-r^3}{3r},&\quad \psi_n(r)=\frac{1}{2}\left(\frac{1}{r^n}-\frac{1}{r^{n-2}}\right) \textrm{  for  }n\geq 2\label{eq:streamn2}.
 \end{align}
In the decomposition above, the first mode is the only one that contributes to the swimming motion (we have $U(t)=\alpha_1(t)$ for all times) and is referred to as the swimming mode, or ``treadmill''. All remaining modes (including the so-called stresslet, $n=2$, characterizing the modification of the bulk stress by the swimmer) correspond to higher order singularities in the far-field flow and do not contribute to the swimming motion.

The dimensionless energetic cost,  $\mathcal{P}$, is computed as \citep{michelin2010c}
\begin{equation}\label{eq:powerdef}
\mathcal{P}=\frac{\mathscr{P}}{12\pi\mu_f a^3\omega^2}=\sum_{n=1}^\infty\gamma_n\mean{\alpha_n^2},
\end{equation}
with 
\begin{equation}
\gamma_1=1\textrm{ and }\gamma_n=\frac{(2n+1)^2}{3n(n+1)} \textrm{ for   } n\geq 2,
\end{equation}
and is equal to the rate of working of the squirmer on the fluid through its boundary actuation or, equivalently, to the total energy loss through viscous dissipation in the fluid domain. In the following, we define $\langle f\rangle=\frac{1}{2\pi}\int_0^{2\pi}f(t)\dd t$ as the time-averaging operator over one stroke period. With this definition, $\mathcal{U}=\sqrt{\mathcal{P}}$ is the typical non-dimensional surface velocity of the swimmer. Following \citet{lighthill1975}, the stroke  swimming efficiency, $\eta$ - or scaled energy cost -, is defined as the ratio of the energetic cost of pulling a rigid sphere with constant velocity $\mean{U}$ and the energetic cost of swimming at the same average velocity, obtained here as \citep{michelin2010c}:
\begin{equation}\label{eq:efficiency}
\eta=\frac{\mean{U}^2}{2\mathcal{P}}=\frac{\mean{\alpha_1}^2}{2\displaystyle\sum_{n=1}^\infty\gamma_n\mean{\alpha_n^2}}\cdot
\end{equation}

\subsubsection{Feeding problem}
To evaluate the amount of nutrient  absorbed at the surface of the organism, the non-dimensional advection-diffusion problem must be solved
\begin{equation}\label{eq:advdiff}
\varepsilon\left(\pard{c}{t}+\ub\cdot\grad c\right)=\nabla^2 c, \textrm{ with   }\varepsilon=\frac{\Pe}{\sqrt{\mathcal{P}}},
\end{equation}
together with the far-field behavior and purely absorbing boundary conditions on the swimmer surface (figure~\ref{fig:squirmer})
\begin{align}
c\rightarrow 0 &\textrm{   for   } r\rightarrow \infty,\label{eq:advdiff_bc1}\\
c=1&\textrm{   for   } r=1.\label{eq:advdiff_bc2}
\end{align}
In equation (\ref{eq:advdiff}) the parameter $\varepsilon=\omega  a^2/\kappa$ can also be understood as the period-based P\'eclet number. The flow field, $\ub$, originates from   the organism stroke and is obtained from the squirming mode amplitudes, $\alpha_n(t)$, using~\eqref{eq:stream}, \eqref{eq:streamn} and \eqref{eq:streamn2}. The feeding performance of the stroke is evaluated using the ratio $J(t)=\Phi(t)/\Phi_0$ quantifying  the net gain in nutrient uptake in comparison with the purely diffusive case ($\Pe=0$). The relative nutrient flux, $J$, is therefore non-dimensional and given by
\begin{equation}\label{eq:Jdef}
J(t)=-\frac{1}{2}\int_{-1}^1\left.\pard{c}{r}\right|_{r=1}\dd\mu.
\end{equation}

\subsubsection{Eulerian vs.~Lagrangian description}

A given periodic stroke, be it swimming or non-swimming, can be mathematically described following two different approaches:
\begin{enumerate}
\item{By prescribing at each instant, a periodic surface velocity on each point fixed in the swimmer frame, $u_\theta^S(\mu,t)$, or equivalently a set of functions $\{\alpha_n(t)\}_n$. We will  refer to this description  in the following as the \emph{Eulerian periodic stroke}.}
\item{By prescribing periodic trajectories, $\xi(\mu_0,t)$, of material surface points labeled by their reference position on the sphere $\mu_0$. We will refer to this description in the following as the \emph{Lagrangian periodic stroke}. The surface velocity and mode amplitudes, $\alpha_n(t)$, can then be obtained from $\xi(\mu_0,t)$ as \citep{michelin2010c}
\begin{align}
&u_\theta(\xi(\mu_0,t),t)=-\frac{1}{\sqrt{1-\xi(\mu_0,t)^2}}\pard{\xi}{t}(\mu_0,t),\\
&\alpha_n(t)=\frac{1}{2}\int_{-1}^1L_n\left[\xi(\mu_0,t)\right]\pardd{\xi}{\mu_0}{t}\dd \mu_0.\label{eq:alphadef}
\end{align}
}
\end{enumerate}

In both descriptions, the flow velocity is periodic and completely determined by the periodic functions $\alpha_n(t)$. However, in the Eulerian formulation, material surface points do  not necessarily have periodic trajectories. Indeed, periodic  Lagrangian strokes only represent a subset of periodic Eulerian  strokes, namely the ones guaranteeing that every surface point comes back to its original position at the end of a full stroke period. Despite its shortcomings regarding the description of material point trajectories, the Eulerian approach has been the most popular for models of swimmers because of its simplicity, and in particular  the possibility to consider steady strokes corresponding to steady surface and flow velocities \citep[][to cite only a few]{ishikawa2006,short2006,doostmohammadi2012,evans2011}.

\subsection{Optimal swimming and optimal feeding}
\label{sec:summary}
For a given amount of energy available to perform a periodic stroke, an organism might have different optimal surface motions depending on the biological function of interest: migration (swimming problem) or nutrient uptake (feeding problem). A priori, those two objectives should lead to different optimal strokes, if anything because the optimal feeding stroke may depend on nutrient  diffusivity  through the value of  $\Pe$ while the swimming problem does not depend on it. 

As emphasized earlier, a periodic stroke can be defined in two different ways, either from an Eulerian point of view (periodic flow field) or from a Lagrangian point of view (periodic material displacement). In our recent contributions, we presented the result of the optimal swimming problem (for both Eulerian and Lagrangian strokes) \citep{michelin2010c} and of the optimal feeding problem in the Eulerian steady framework only \citep{michelin2011}. A brief summary of these results is first presented here.

We start by remarking  that, for the swimming problem, Eulerian optimal strokes are necessarily steady and each mode,  $\alpha_n$, is independent of  time.  This is a direct consequence of the absence of history effect in the swimming problem: the swimming velocity and the energetic cost only depend on the instantaneous surface velocity. The optimal Eulerian stroke is then obtained by choosing the surface velocity distribution maximizing {instantaneously} the efficiency $\eta$. From~\eqref{eq:efficiency} we see that the Eulerian optimal swimming stroke is simply obtained by putting all the energy into the swimming mode, namely $\alpha_n(t)=\delta_{n,1}$. The resulting treadmill swimmer, with an efficiency $\eta_\textrm{max}=50\%$, is therefore the overall  optimal for locomotion \citep{leshansky2007,michelin2010c}. 

In the case of the feeding problem, the presence of a time-derivative in the advection-diffusion equation introduces history effects, and the optimal Eulerian feeding stroke is therefore not necessarily steady. Focusing on the simplified problem of steady strokes, \citet{michelin2011} showed using adjoint-based optimization that the optimal  steady feeding stroke is essentially the same as the optimal steady swimming stroke, a result which, surprisingly, remains true for all P\'eclet number. 

That result was not obvious  {a priori}. The value of the mean feeding rate of the organism for a given stroke is a  strong function of  the  diffusivity of the nutrient whose distribution around the organism is qualitatively different in the diffusive and advective regimes \citep{magar2003,michelin2011}. The optimal feeding rate, $\mean{J}_\textrm{opt}$, depends strongly on $\Pe$, but the stroke to achieve this optimal value does not. This result is important biologically as it implies that, for 
   a given organism, a unique optimal stroke maximizes the nutrient uptake regardless of the details of its diffusive transport. For all $\Pe$, and in the steady Eulerian framework, {maximizing feeding and maximizing  swimming are  therefore equivalent problems}.

Although simpler conceptually and mathematically, the Eulerian framework is not appropriate to describe periodic deformations of a material surface, such as, for example, the  strokes of ciliated cells. To impose periodicity of the surface motion, it is  necessary to turn to the Lagrangian approach {and to consider the unsteady swimming and feeding problems}. \citet{michelin2010c} showed numerically that the optimal Lagrangian swimming stroke could be decomposed into two different parts: an effective stroke, dominated by the swimming mode, $\alpha_1$, and producing a forward velocity, and a recovery stroke during which material points (e.g. cilia tips) are brought back to their original position with front-like dynamics to minimize their (negative) impact on the swimming velocity. This front, or wave, is reminiscent of  metachronal waves observed on ciliated organisms \citep{brennen1977} and results from a small phase-shift in the motion of neighboring surface points leading to global symmetry-breaking at the whole-organism level.  When the squirmer model is used to represent a ciliate, the cilia length constrains the maximum displacement of the surface and therefore limits the ability of the swimmer to not only approach the optimal Eulerian stroke {(treadmill)} during the effective stroke but also to  reduce the impact of the recovery stroke on the swimming motion. Using a constrained optimization algorithm, the direct relationship between swimming efficiency and surface displacement amplitude was obtained, and \citet{michelin2010c} showed that the optimal  efficiency of $50\%$ could be reached asymptotically.

The optimization of the Lagrangian feeding stroke  however remains at this point an open question; it is the focus of the present paper. The analysis of the nutrient uptake  is first  addressed analytically at small $\Pe$. The general unsteady feeding problem is then considered  numerically  before turning to its optimization.

\section{Unsteady feeding at low $\Pe$: Asymptotics, scalings, and optimum}
\label{sec:asymptotics}
In this section we focus on the feeding problem in the asymptotic limit of dominant diffusion  ($\Pe\ll 1$). For a given stroke, this is equivalent to the asymptotic analysis of the advection-diffusion problem in the limit   $\varepsilon=\Pe/\sqrt{\mathcal{P}}\ll 1$.

\subsection{Steady and unsteady boundary layers}
For a steady velocity field, finding the asymptotic expansion of the scalar concentration, $c$, and surface flux, $J$, in the limit $\varepsilon\ll 1$ corresponds to a variation on the classical mass transfer problem near a sedimenting sphere \citep{acrivos1962,magar2003,michelin2011}. It is based on matching two different solutions for the scalar field $c$:  near the surface of the sphere, diffusive effects are dominant, and advection only appears as higher order corrections, while  in the far-field, a balance of both advection and diffusion leads to the proper decay of $c$. 

In the case of an unsteady velocity field, both terms on the left hand-side of~\eqref{eq:advdiff} do not have the same scaling in the far-field. As a result the decay of the concentration field at infinity is not the same whether one considers the time-average of $c$ or its fluctuations around the mean, and a double boundary layer problem must be considered:

\begin{itemize}
\item{in the near field, $r=O(1)$, diffusion dominates and the absorbing boundary condition ($c=1$) at the surface of the swimmer  is satisfied;}
\item{in the unsteady boundary layer (UBL), $R=\varepsilon^{1/2} r=O(1)$, a balance between diffusive effects and rate of change of the local concentration ensures the proper far-field decay for the time-dependent fluctuations of the concentration field $\mathcal{C}(R,\mu,t)=c(r,\mu,t)$;}
\item{in the steady boundary layer (SBL), $\rho=\varepsilon\, r=O(1)$, a balance between advection by the steady velocity field and diffusion ensures the far-field decay of the time-average concentration $\mathscr{C}_0(\rho,\mu)=\mean{c}(r,\mu)$.}
\end{itemize}

\subsection{Asymptotic problem formulation}
Decomposing the mode amplitudes, $\alpha_n(t)$, as well as the concentration field, $c$, and feeding rate, $J(t)$, into their Fourier components, we write
\begin{equation}
\alpha_n(t)=\sum_{p=-\infty}^\infty\alpha_{n,p}\ee^{\ci pt},\quad
c=\sum_{p=-\infty}^\infty c_p(r,\mu)\ee^{\ci pt},\quad J(t)=\sum_{p=-\infty}^\infty J_p\ee^{\ci pt}.
\end{equation}
The advection-diffusion equation becomes then
\begin{eqnarray}
&\textrm{in the near field,}\qquad &D\cdot c_p=\varepsilon\left(\ci pc_p+\sum_{n=1}^\infty\sum_{q=-\infty}^\infty\alpha_{n,q}l_n\cdot c_{p-q}\right),\label{eq:advdiff_nearfield}
\\
&\textrm{in the UBL,}\qquad &\mathcal{D}\cdot \mathcal{C}_p=\ci p\,\mathcal{C}_p+\varepsilon^{1/2}\sum_{q=-\infty}^\infty\alpha_{1,q}\mathcal{L}_1\cdot \mathcal{C}_{p-q} +O(\varepsilon^{3/2}),\label{eq:advdiff_UBL}\\
&\textrm{in the SBL,}\qquad&\mathscr{D}\cdot \mathscr{C}_0=\alpha_{1,0}\,\mathscr{L}_1\cdot \mathscr{C}_0 +O(\varepsilon^2).\label{eq:advdiff_SBL}
\end{eqnarray}
In~\eqref{eq:advdiff_nearfield}--\eqref{eq:advdiff_SBL}, the following linear operators have been defined
\begin{align}
D&=\frac{1}{r^2}\left[\pard{}{r}\left(r^2\pard{}{r}\right)+\pard{}{\mu}\left((1-\mu^2)\pard{}{\mu}\right)\right],\label{eq:diff_op}\\
l_1&=-\left(1-\frac{1}{r^3}\right)\mu\pard{}{r}-\frac{1-\mu^2}{r}\left(1+\frac{1}{2r^3}\right)\pard{}{\mu},\label{eq:advl1}\\
l_n&=\frac{2n+1}{2}\left[\left(\frac{1}{r^{n+2}}-\frac{1}{r^n}\right)L_n(\mu)\pard{}{r}-\frac{(1-\mu^2)L_n'(\mu)}{n(n+1)}\left(\frac{n}{r^{n+3}}-\frac{n-2}{r^{n+1}}\right)\pard{}{\mu}\right],\label{eq:advln}\\
\mathcal{L}_1&=-\mu\pard{}{R}-\frac{(1-\mu^2)}{R}\pard{}{\mu},\label{eq:adv_op_bl}
\end{align}
and $\mathcal{D}$ (resp. $\mathscr{D}$) is identical to $D$ in~\eqref{eq:diff_op} after replacing $r$ by $R$ (resp. $\rho$), and $\mathscr{L}_1$ is defined as $\mathcal{L}_1$ after replacing $R$ by $\rho$. The following boundary conditions must also be satisfied:
\begin{align}
\forall p, &\quad c_p(r=1)=\delta_{p,0},\label{eq:bc_surface}\\
\forall p\neq 0,&\quad \mathcal{C}_p(R\rightarrow\infty)=0 ,\\
&\quad\mathscr{C}_0(\rho\rightarrow \infty)=0.
\end{align}

\subsection{Matched Asymptotic Expansion}

A regular series expansion in $\varepsilon^{1/2}$ of $c_p$, $\mathcal{C}_p$ and $\mathscr{C}_0$ is then performed up to $O(\varepsilon^{3/2})$. We write
\begin{equation}
c_p=\sum_{q=0}^3\varepsilon^{q/2}c_p^q+O(\varepsilon^2),\quad \mathcal{C}_p=\sum_{q=0}^3\varepsilon^{q/2}\mathcal{C}_p^q+O(\varepsilon^2),\quad\mathscr{C}_0=\sum_{q=0}^3\varepsilon^{q/2}\mathscr{C}_0^q+O(\varepsilon^2).
\end{equation}
At each order, $c_p$ and $\mathcal{C}_p$ are to be matched for $r\rightarrow \infty$ and $R\rightarrow 0$, while $\mathcal{C}_0$ and $\mathscr{C}_0$ are to be matched in the limit $R\rightarrow \infty$ and $\rho\rightarrow 0$. 

Here, the non-homogeneous forcing \eqref{eq:bc_surface} only acts on the steady-state component of the concentration field, and is transmitted to the time-dependent components by advection. Therefore, from the scalings of the different terms in equations \eqref{eq:advdiff_nearfield}-\eqref{eq:advdiff_UBL},
\begin{equation}
\forall p\neq 0, \quad c_p=O(\varepsilon c_0)\quad \textrm{ and} \quad \mathcal{C}_p=O(\varepsilon^{1/2}\mathcal{C}_0). \label{eq:scalingMAE}
\end{equation}

\subsubsection{Order $O(1)$}
At this order, advection is neglected and the solution is simply the steady diffusive solution $c_p^0=\delta_{p,0}/r$, which satisfies both near-field and far-field boundary conditions. Therefore, $\mathcal{C}_p^0=\mathscr{C}_0^0=0$ for all $p$. The resulting feeding rate is 
\begin{equation}
J_p=\delta_{p,0}+O(\varepsilon^{1/2}).
\end{equation}

\subsubsection{Order $O(\varepsilon^{1/2})$}
Using~\eqref{eq:scalingMAE}, $c_p^1=0$ and $\mathcal{C}_p^1=0$ for all $p\neq 0$. The steady components $c_0^1$, $\mathcal{C}_0^1$, and $\mathscr{C}_0$ satisfy
\begin{align}
D\cdot c_0^1&=0,\\
\mathcal{D}\cdot\mathcal{C}_0^1&=0,\\
\mathscr{D}\cdot\mathscr{C}_0^1&=\alpha_{1,0}\mathscr{L}_1\cdot\mathscr{C}_0^1.
\end{align}
Solving these equations and matching $c_p$, $\mathcal{C}_p$ and $\mathscr{C}_0$ up to $O(\varepsilon^{1/2})$ leads to 
\begin{equation}
c_p^1=0,\quad
\mathcal{C}_p^1=\frac{\delta_{p,0}}{R},\quad \mathscr{C}_0^1=0.\label{eq:res_ordre1}
\end{equation}
and the resulting feeding rate remains unmodified at this order.

\subsubsection{Order $O(\varepsilon)$}
Next, the advection diffusion equation is expanded up to $O(\varepsilon)$ in each region.
\begin{itemize}
\item{\emph{In the near field, $r=O(1)$}:
\begin{equation}
D\cdot c_p^2=\sum_{n=1}^\infty\alpha_{n,p}\,l_n\cdot c_0^0=\frac{\mu\alpha_{1,p}}{r^2}\left(1-\frac{1}{r^3}\right)-\sum_{n=2}^\infty\frac{(2n+1)\alpha_{n,p}L_n(\mu)}{2r^2}\left(\frac{1}{r^{n+2}}-\frac{1}{r^n}\right), \label{eq:modif1}
\end{equation}
whose general solution satisfying the near-field boundary condition, $c_p^2(r=1)=0$, is obtained as
\begin{align}
c_p^2(r,\mu)=\alpha_{1,p}\mu&\left(\frac{3}{4r^2}-\frac{1}{2}-\frac{1}{4r^3}\right)+\sum_{n=1}^\infty \gamma_{n,p}L_n(\mu)\left(\frac{1}{r^{n+1}}-r^n\right)\nonumber\\
&-\sum_{n=2}^\infty\frac{(2n+1)\alpha_{n,p}L_n(\mu)}{4}\left(\frac{1}{(n+1)r^{n+2}}+\frac{1}{nr^n}-\frac{2n+1}{n(n+1)r^{n+1}}\right),\label{eq:modif2}
\end{align}
where $\gamma_{n,p}$ are constants to be determined after matching with the UBL solution.
}
\item{\emph{In the unsteady boundary layer, $R=O(1)$}:
\begin{equation}
\mathcal{D}\cdot\mathcal{C}_p^2-\ci p\,\mathcal{C}_p^2=\alpha_{1,p}\mathcal{L}_1\cdot\mathcal{C}_0^1=\frac{\mu\alpha_{1,p}}{R^2},
\end{equation}
whose general solution compatible with the boundary condition at infinity (for $p\neq 0$) is
\begin{align}
\mathcal{C}_0^2&=\frac{\gamma_{0,0}'}{R}-\gamma''_{0,0}+\mu\left(\frac{\gamma'_{1,0}}{R^2}-\gamma''_{1,0}R-\frac{\alpha_{1,0}}{2}\right),\\
\mathcal{C}_p^2&=\frac{\gamma'_{0,p}}{R}\ee^{-R\sqrt{\ci p}}+\mu\left[\frac{\ci\alpha_{1,p}}{pR^2}+\gamma'_{1,p}\left(\frac{1}{R^2}+\frac{\sqrt{\ci p}}{R}\right)\ee^{-R\sqrt{\ci p}}\right],\textrm{    for    } p\neq 0.
\end{align}

}
\item{\emph{In the steady boundary layer, $\rho=O(1)$,} the general solution of~\eqref{eq:advdiff_SBL} is obtained as \citep{acrivos1962}
\begin{equation}\label{eq:gensolSBL}
\mathscr{C}_0^2=\frac{1}{\rho}\,\mathrm{exp}\left(-\displaystyle\frac{\alpha_{1,0}(1+\mu)\rho}{2}\right)\sum_{q=0}^\infty K_q^2L_q(\mu)\left(\sum_{m=0}^q\frac{(q+m)!}{(\alpha_{1,0}\rho)^m\,m!(q-m)!}\right),
\end{equation}
}
\end{itemize}
where the $K_q^2$ are constants to be determined in the matching process. Matching $c_p$, $\mathcal{C}_p$, and $\mathscr{C}_0$, up to $O(\varepsilon)$ leads to 
\begin{align}
c_p^2&=\frac{\alpha_{1,0}}{2}\left(\frac{1}{r}-1\right)\delta_{p,0}+\mu\alpha_{1,p}\left(-\frac{1}{2}+\frac{3}{4r^2}-\frac{1}{4r^3}\right)\nonumber\\
&-\sum_{n=2}^\infty\frac{(2n+1)\alpha_{n,p}L_n(\mu)}{4}\left(\frac{1}{(n+1)r^{n+2}}+\frac{1}{nr^n}-\frac{2n+1}{n(n+1)r^{n+1}}\right),\label{eq:res_ordre2}\\
\mathcal{C}_0^2&=-\frac{\alpha_{1,0}(1+\mu)}{2},\quad \mathcal{C}_p^2=\frac{\ci\alpha_{1,p}\mu}{p\,R^2}\left[1-\left(1+R\sqrt{\ci p}\right)\ee^{-R\sqrt{\ci p}}\right]\textrm{    for    } p\neq 0,\label{eq:res_ordre2b}\\
\mathscr{C}_0^2&=\frac{1}{\rho}\,\mathrm{exp}\left(-\displaystyle\frac{\alpha_{1,0}(1+\mu)\rho}{2}\right),\label{eq:res_ordre2bb}
\end{align}
and the resulting feeding rate expansion is 
\begin{equation}
J_p=\delta_{p,0}\left(1+\frac{\varepsilon\alpha_{1,0}}{2}\right)+O(\varepsilon^{3/2}).
\end{equation}

Up to this order, we see that the results of the classical low-$\Pe$ asymptotic expansion for a steady velocity field are recovered and the mean feeding rate only depends on the average swimming velocity. In order to capture the leading order unsteady contribution to the feeding problem, the expansion must be carried out to the next order.

\subsubsection{Order $O(\varepsilon^{3/2})$}
From~\eqref{eq:Jdef}, we see that only the computation of the azimuthal average, $\tilde{c}_p^3$, of the $p$-th Fourier component of the concentration field 
\begin{equation}
\tilde{c}_p^3(r)=\frac{1}{2}\int_{-1}^1c_p^3(r,\mu)\dd \mu,
\end{equation}
is necessary in order to compute the $O(\varepsilon^{3/2})$ correction to the nutrient uptake.

\begin{itemize}
\item{\emph{In the near-field}, taking the azimuthal average of~\eqref{eq:advdiff_nearfield} and using~\eqref{eq:res_ordre1}, we have
\begin{equation}
\frac{1}{r^2}\totd{}{r}\left(r^2\totd{\tilde{c}_p^3}{r}\right)=0,
\end{equation}
whose general solution satisfying the boundary condition on the sphere is $\tilde{c}_p^3=a_p(1-1/r)$, where $a_p$ is a constant to be determined by matching with the UBL solution.
}
\item{\emph{In the unsteady boundary layer}, taking the azimuthal average of~\eqref{eq:advdiff_UBL} and using~\eqref{eq:res_ordre2b}, we get
\begin{align}
\frac{1}{R^2}\totd{}{R}\left(R^2\totd{\tilde{\mathcal{C}}_p^3}{R}\right)-\ci p \tilde{\mathcal{C}}_p^3&=\frac{1}{2}\sum_{q=-\infty}^\infty\alpha_{1,p-q}\int_{-1}^1\mathcal{L}_1\cdot\mathcal{C}_p^2\dd\mu\nonumber\\
&=\frac{1}{3R}\sum_{q=-\infty}^\infty\alpha_{1,p-q}\alpha_{1,q}\ee^{-R\sqrt{\ci q}}.\label{eq:modif4}
\end{align}

This equation can be solved explicitly for $\tilde{C}_p^3$ using the far-field boundary condition for the non-constant Fourier components and we get
\begin{align}
\tilde{\mathcal{C}}_0^3=\frac{\alpha_{1,0}^2R}{6}+\frac{\tilde{a}_0}{R}&+\tilde{b}_0-\sum_{m=1}^\infty\frac{2|\alpha_{1,m}|^2}{3mR}\ee^{-R\sqrt{m/2}}\sin\left(R\sqrt{\frac{m}{2}}\right),\\
\tilde{\mathcal{C}}_p^3=\frac{\tilde{a}_p}{R}\ee^{-R\sqrt{\ci p}}&+\frac{\alpha_{1,0}\alpha_{1,p}}{3}\left(\frac{\ci}{p R}-\frac{\ee^{-R\sqrt{\ci p}}}{2\sqrt{\ci p}}\right)+\sum_{m=1}^{p-1}\left(\frac{\ci\alpha_{1,m}\alpha_{1,p-m}}{3mR}\right)\ee^{-R\sqrt{\ci(p-m)}}\nonumber\\
&+\sum_{m=p+1}^\infty\left(\frac{\ci\alpha_{1,m}\overline{\alpha_{1,m-p}}}{3R}\right)\left(\frac{\ee^{-R\sqrt{\ci(p-m)}}}{m}+\frac{\ee^{-R\sqrt{\ci m}}}{p-m}\right) \textrm{ for     }p\geq 1,
\end{align}
with $\tilde{\mathcal{C}}_p^3$   defined for $p\leq -1$ using $\tilde{\mathcal{C}}_{-p}=\overline{\tilde{\mathcal{C}}_p}$. 
}
\item{\emph{In the steady boundary layer}, the equation for $\mathscr{C}_0^3$ is identical to that at the previous order and the general solution takes the same form, see \eqref{eq:gensolSBL}.}
\end{itemize}

By matching $\tilde{c}_p$, $\tilde{\mathcal{C}}_p$, and $\tilde{\mathscr{C}}_0$ up to $O(\varepsilon^{3/2})$, the values of $\tilde{b}_0$, $\tilde{a}_p$, and $a_p$ can then be determined, and one obtains:
\begin{align}
\langle J\rangle&=1+\frac{\varepsilon\alpha_{1,0}}{2}+\varepsilon^{3/2}\frac{\sqrt{2}}{3}\sum_{m=1}^\infty\frac{|\alpha_{1,m}|^2}{\sqrt{m}}+O(\varepsilon^2)\label{eq:predict_mean}\\
J(t)&-\langle J\rangle=\varepsilon^{3/2}\left(-\ci\sqrt{\ci}\right)\sum_{p\neq 0}\tilde{J}_p\ee^{\ci pt}+O(\varepsilon^2),\label{eq:predict_fluct},
\end{align}
with
\begin{align}
\tilde{J}_p=\left[\frac{\alpha_{1,0}\alpha_{1,p}}{2\sqrt{p}}\right.&+\sum_{m=1}^{p-1}\frac{\alpha_{1,m}\alpha_{1,p-m}}{3m}\left(\sqrt{p}-\sqrt{p-m}\right)\nonumber\\
&\left.+\sum_{m\geq p+1}\frac{\alpha_{1,m}\overline{\alpha_{1,m-p}}}{3m(m-p)}\left(m^{3/2}-p^{3/2}-\ci(m-p)^{3/2}\right)\right]\label{eq:predict_fluct2}.
\end{align}
For a given stroke, the limit $\varepsilon\ll 1$ is equivalent to $\Pe\ll 1 $ and the asymptotic expansion in terms of the P\'eclet number, $\Pe$, can be obtained by substitution of $\varepsilon=\Pe/\sqrt{\mathcal{P}}$ in~\eqref{eq:predict_mean}--\eqref{eq:predict_fluct2}.

\subsection{Discussion}
The asymptotic analysis obtained in~\eqref{eq:predict_mean}--\eqref{eq:predict_fluct2}  provides some important physical insight into the relationship between the swimming motion and the nutrient uptake on the surface of the swimmer. As for the steady case, the leading order advective correction to the feeding rate is linear in $\Pe$ and only depends on the average velocity of the organism \citep{acrivos1962,magar2003}. At this order in $\Pe$, there is a direct correlation between swimming and feeding and only the mean feeding rate is modified, fluctuations in time being negligible (higher order).

The next order correction marks a fundamental difference between the steady and unsteady problems: in the steady case, all squirming modes contribute to the next correction at order  $\Pe^2$ \citep{michelin2011}. Instead, in the unsteady feeding problem, a new correction to $J(t)$ (both its mean value in time and fluctuations) appears at order  $\Pe^{3/2}$, which  depends solely on the swimming velocity of the organism (through all the  Fourier components, $\alpha_{1,m}$, of the swimming velocity, $\alpha_1(t)$, with no other squirming modes), and dominates the contribution of non-swimming modes that will only enter at order $O(\Pe^2)$. For all time-periodic strokes, the instantaneous feeding rate is therefore completely determined up to $O(\Pe^{3/2})$ by the characteristics of the swimming velocity of the organism.

This result has a major consequence for strokes that swim instantaneously ($U(t)\neq 0$) but do not swim on average ($\mean{U}=0$). In this case, the leading-order improvement to the feeding rate is solely governed by the zero-mean fluctuations of $U(t)$. Non-swimming modes only contribute to higher order corrections, even if they have non-zero time averages. Consequently, for an organism that does not have a net swimming motion (e.g. a time-reversible swimmer), an instantaneous zero-mean swimming motion still presents a feeding advantage over stirring strokes where the cell stays in the same position at each instant ($U(t)=0$).

Our asymptotic expansion also provides some information on the relative phase of swimming and feeding. For an unsteady swimming velocity, $U(t)$, with a single dominant Fourier component, the instantaneous feeding rate has a $\pi/4$ delay on the swimming velocity (since $-\ci\sqrt\ci=\ee^{-\ci\pi/4}$ in
\ref{eq:predict_fluct}). A maximum in the feeding rate is therefore expected to take place after the peak swimming velocity, with a delay of 1/8$^\textrm{th}$ of a period.

Note that  the total nutrient flux is fully determined by the body velocity $U(t)$ up to $O(\Pe^{3/2})$ . Whether the organism is swimming (force-free) or is an actuated rigid sphere (forced motion) does not actually come into play here.  All the conclusions above are therefore valid for non-buoyant swimmers, but also for oscillating rigid spheres in Stokes flow, for which the present results represent a generalization of classical steady mass transfer results \citep{acrivos1962} to unsteady motions (see Appendix \ref{sec:acrivos_unsteady} for more details).

In summary, our analytical results show that for low $\Pe$, feeding is completely determined by swimming for any periodic stroke. Optimization of the feeding rate for a fixed amount of available energy is therefore equivalent in this limit to maximizing the swimming velocity under the same constraint, namely the swimming efficiency optimization problem.  At low P\'eclet number, the Lagrangian optimal swimming and optimal feeding strokes are therefore identical, which confirms the result obtained in the steady framework by \citet{michelin2011}. In addition, similarly to the result for swimming, we get the result that at low P\'eclet number the optimal unsteady feeding problem is actually steady. This can be seen from  \eqref{eq:predict_mean} where the steady Fourier mode, $\alpha_{1,0}$, carries a higher  weight than the other Fourier components compared to their relative importance in the  rate of working.

\section{Unsteady feeding at finite $\Pe$: simulations}
\label{sec:unsteady_simulations}
To confirm the low-$\Pe$ results obtained analytically, we now turn to 
characterizing the  feeding performance of different strokes for intermediate and large $\Pe$. Eulerian periodic strokes are determined by prescribing $\alpha_n(t)$ for all $n$, while Lagrangian periodic strokes are described by giving the trajectories of material points $\theta=\vartheta(\theta_0,t)$ where $\theta$ is the current position of the material point and $\theta_0$ its mean position. Alternatively, those strokes will be defined by $\mu=\xi(\mu_0,t)$, with $\mu=\cos\theta$. For illustration we consider  three particular swimming and non-swimming Lagrangian periodic strokes:
\begin{enumerate}
\item{\emph{Stroke A} is the numerical optimal swimmer identified in \citet{michelin2010c} which has swimming efficiency $\eta\approx20\%$;}
\item{\emph{Stroke B} is a less efficient swimmer obtained using surface deformations in the form of a simple progressive wave:
\begin{equation}
\xi(\mu_0,t)=\mu_0+A(1-\mu_0^2)\cos(k\mu_0-t),
\end{equation}
with $A=1/3$ and $k=1$;}
\item{\emph{Stroke C} takes the same form as stroke B but with $A=1/3$ and $k=0$. Stroke C represents a time-reversible (or ``reciprocal'') deformation, and therefore has no net swimming motion, $\mean{U}=0$.}
\end{enumerate}

\begin{figure}
\begin{center}
\includegraphics[width=.95\textwidth]{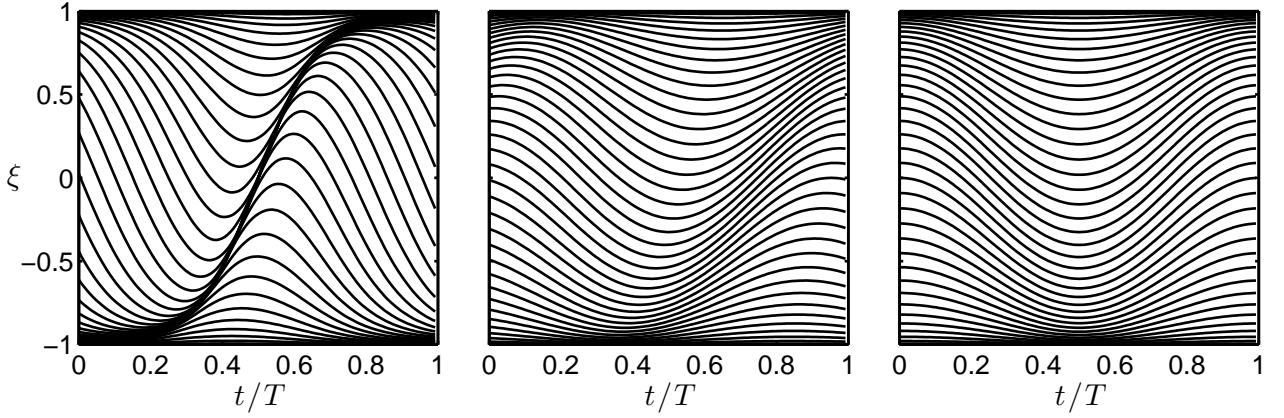}
\caption{Trajectories, $\xi(t)$, of surface material points for stroke A (left), stroke B (center), and stroke C (right). The corresponding time-averaged swimming velocity, $\langle U\rangle$, is equal to $0.33$, $0.03$, and $0$  respectively. The swimming and feeding performances of the three strokes are summarized in Table \ref{tab:strAB}.}\label{fig:trajectories}
\end{center}
\end{figure}

All three strokes display non-zero instantaneous swimming, but only strokes 
A and B show swimming on average.  Stroke C differs thus from purely stirring strokes for which the organism is strictly still at each instant. The trajectories of material surface points are shown for strokes A, B and C in figure \ref{fig:trajectories}. Mathematically, from the knowledge of $\xi(\mu_0,t)$, the mode amplitudes $\alpha_n(t)$ are obtained using~\eqref{eq:alphadef}.

\subsection{Numerical solution of the advection diffusion problem}
For a given set of mode amplitudes, $\{\alpha_n(t)\}$, the advection-diffusion equation in~\eqref{eq:advdiff} is solved spectrally in time for each azimuthal component of the concentration field
\begin{equation}
c(r,\mu,t)=\sum_{p=0}^\infty c^*_p(r,t)L_p(\mu)=\sum_{k=-\infty}^\infty\sum_{p=0}^\infty c_p^k(r)L_p(\mu)\ee^{\ci kt}.
\end{equation}
The functions $c_p^k(r)$ satisfy therefore the following systems of ordinary differential equations for $p\geq 0$ and $-\infty<k<\infty$:
\begin{align}
\left[\frac{1}{\Pe}\left(\totd{^2}{r^2}+\frac{2}{r}\totd{}{r}-\frac{p(p+1)}{r^2}\right)-\ci k\right]&c_p^k=\nonumber\\
&\sum_{m=0}^\infty\sum_{n=1}^\infty\sum_{l=-\infty}^\infty\frac{\alpha_n^{k-l}}{r^2}\left(A_{mnp}\psi_n\totd{}{r}+B_{mnp}\totd{\psi_n}{r}\right)c_m^l\label{eq:advdiff_discrete},
\end{align}
with boundary conditions
\begin{align}
c_p^k(r=1)&=\delta_{p,0}\delta_{k,0},\\
c_p^k(r\rightarrow\infty)&=0.\label{eq:discretebc2}
\end{align}
In~\eqref{eq:advdiff_discrete}, $A_{mnp}$ and $B_{mnp}$ are third-order scalar tensors defined in Appendix~\ref{sec:AB}. Equations~\eqref{eq:advdiff_discrete}--\eqref{eq:discretebc2} are discretized on an exponentially-stretched grid in $r$ to concentrate points near the surface of the swimmer \citep[see][for more details]{michelin2011}, and the solution $\left\{c_p^k(r_j)\right\}_{(j,k,p)}$ is then found iteratively. In typical simulations, the resolution used was $N_r=120$ points for the $r$-grid, $N_\mu=40$--$100$ Legendre polynomials for the azimuthal dependence, $N_t=16$--$128$ points in time, and $N_\alpha=2$--$10$ squirming modes to describe the swimming stroke. 

Alternatively, the advection-diffusion equation can be marched in time for each azimuthal component, $c_p^*(r,t)$, using an explicit time-stepping scheme for the advective terms and Crank Nicholson for the diffusion term. In the following, the advection-diffusion equation is solved spectrally in time except for strokes that do not swim on average (e.g. stroke C) for which the iterative algorithm does not converge properly or fast enough, and the time-marching approach is used in that case.

\begin{figure}
\begin{center}
\includegraphics[width=10cm]{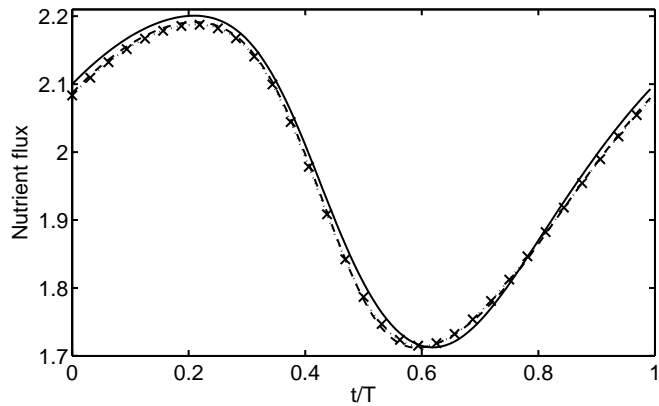}
\caption{Instantaneous nutrient flux at $\Pe=5$ for the optimal swimmer (stroke A) using an increasing number of squirming modes to numerically describe the stroke in the advection-diffusion solver: $n_\alpha=1$ (solid), $n_\alpha=2$ (dashed), $n_\alpha=4$ (dotted) and $n_\alpha=8$ (crosses). The error made on the average nutrient flux over a period is respectively $0.5\%$, $0.03\%$, $0.01\%$ and $0.002\%$.}\label{fig:nalphac}
\end{center}
\end{figure}

Computationally, it is observed that the instantaneous nutrient flux converges rapidly with the number of squirming modes used to represent the swimming stroke, as shown in figure \ref{fig:nalphac}. The convergence is even faster for the average nutrient flux: describing stroke A with only the first two squirming modes significantly speeds up the computations while introducing an error smaller than $0.05\%$ on the average feeding rate. Similar numerical tests performed on less efficient swimmers than stroke A (that is, swimming strokes for which mode 1 is not dominant) did not modify this observation significantly, and restricting the computation to only 2 or 3 squirming modes typically introduces an error smaller than $0.2\%$. This rapid convergence of the mean and fluctuating feeding rate is yet another indication that the  swimming motion controls  the feeding ability of the organism and higher-order modes only act as a small correction to the average feeding rate.

\subsection{Impact of the swimming stroke on the feeding performance}
\begin{figure}
\begin{center}
\includegraphics[width=.91\textwidth]{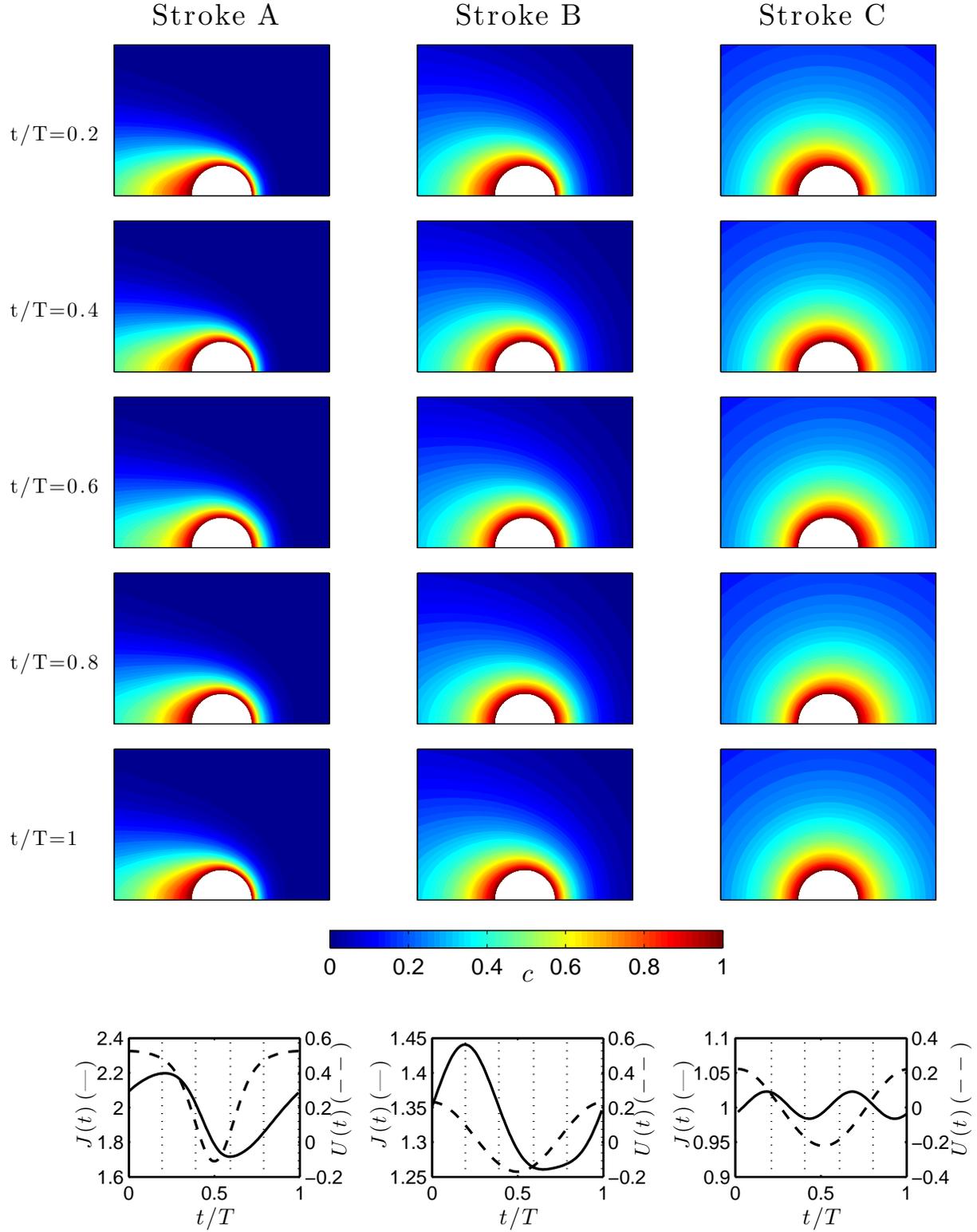}
\caption{Top: Nutrient concentration around the organism at $\Pe=5$ for stroke A (left), stroke B (center) and stroke C (right).  Bottom: Evolution in time of the feeding rate (solid) and swimming velocity (dashed). The dotted lines on the bottom figures indicate the time corresponding to each of the five top snapshots (ordered from top to bottom and left to right).}\label{fig:moviePe5strAB}
\end{center}
\end{figure}

\begin{figure}
\begin{center}
\includegraphics[width=.92\textwidth]{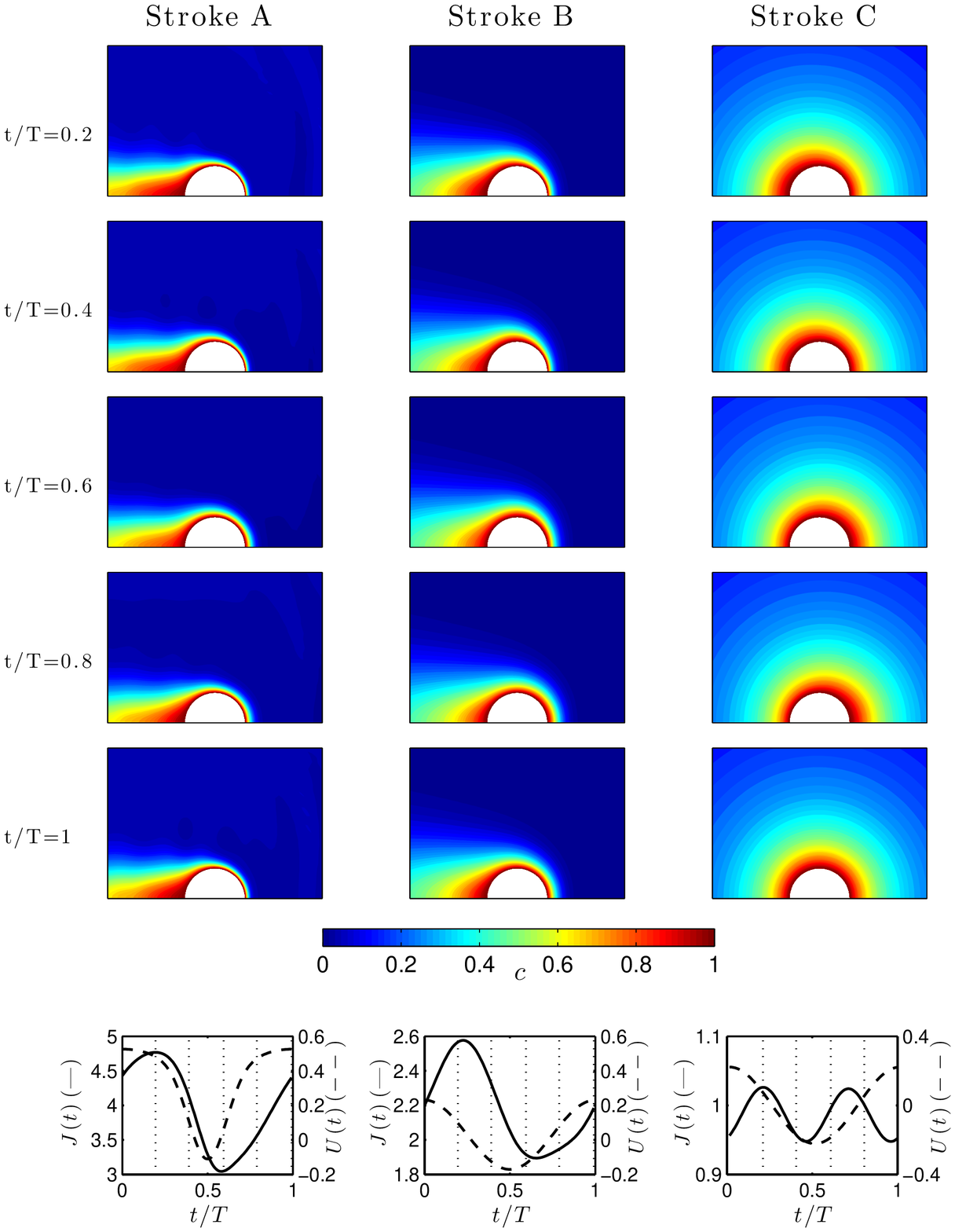}
\caption{Same as figure~\ref{fig:moviePe5strAB} with $\Pe=30$.}\label{fig:moviePe30strAB}
\end{center}
\end{figure}

Figures \ref{fig:moviePe5strAB} and \ref{fig:moviePe30strAB} show the concentration field around the squirmer for five successive and equispaced instants of a full period, for $\Pe=5$ (figure \ref{fig:moviePe5strAB}) and $\Pe=30$ (figure \ref{fig:moviePe30strAB}), and for the three different strokes. For strokes A and B, at lower P\'eclet number, the nutrient concentration field only shows a weak front--back anisotropy as diffusion dominate over advection, confirming the observations on steady strokes of \citet{magar2003} and \citet{michelin2011}. As $\Pe$ is increased, sharper concentration gradients can be seen on the front of the squirmer. This results in an increased average feeding rate for increasing $\Pe$ as was observed for steady strokes \citep{michelin2011}.  The main difference with the steady results is that in the unsteady scenario,  the velocity of the squirmer changes (and possibly reverses sign) inducing a fluctuation in this front-back anisotropy and in the boundary layer thickness. 
For stroke C, which does not swim on average, the nutrient concentration field shows a strong isotropy, even at larger $\Pe$, with much weaker concentration gradients resulting in a very weak modification of the nutrient uptake $\mean{J}$.

\begin{table}
\begin{center}
\begin{tabular}{ccccc}
Stroke &$\quad\mean{U}\quad$ & $\quad \eta\quad$ & $\quad\mean{J}\quad (\Pe=5)\quad$ & $\quad\mean{J}\quad(\Pe=30)\quad$ \\
\\
A &$0.33$ & $22\%$ & $1.97$ & $3.98$\\
B &$0.030$ & $1.3\%$ & $1.33$ & $2.19$ \\
C & $0$ & $0\%$ & $1.00 $ & $0.99 $\\
\\
\end{tabular}
\caption{Swimming and feeding performance of strokes A, B, and C.}\label{tab:strAB}
\end{center}
\end{table}

Comparing the results obtained for the different strokes in Table~\ref{tab:strAB}, we see that stroke A is clearly more efficient than strokes B and C from a feeding point of view, and stroke A also corresponds to a ``better" swimmer. This is consistent with the increase of the feeding rate with the instantaneous swimming velocity that enables the formation of sharp concentration gradients in front of the squirmer. For stroke C, the periodic reversal of the swimming velocity over the  period, and the absence of net displacement, results in the impossibility to maintain sharp concentration gradients at the front of the body and to swim toward regions with richer nutrient content, reducing its feeding ability significantly. 

Looking at the temporal variations of the swimming velocity and feeding rate throughout the stroke period (bottom frames of figures~\ref{fig:moviePe5strAB} and \ref{fig:moviePe30strAB}), a phase delay between the former and the latter is clearly identified for stroke A and B, and for all $\Pe$ considered. For stroke C, a similar delay is observed between the peaks in velocity magnitude (positive or negative) and the peaks in feeding rate: for this stroke, the feeding rate frequency is twice that of the swimming velocity because of the exact symmetry between the two half stroke periods. The presence of this time delay in all strokes is consistent with the results of the low-$\Pe$ asymptotic analysis in \S\ref{sec:asymptotics} and can be interpreted as the time necessary for the concentration gradient (and possibly boundary layer) to reestablish at the front of the cell when its velocity starts  increasing again.

\subsection{Impact of the P\'eclet number on the feeding performance}
\begin{figure}
\begin{center}
\includegraphics[width=.95\textwidth]{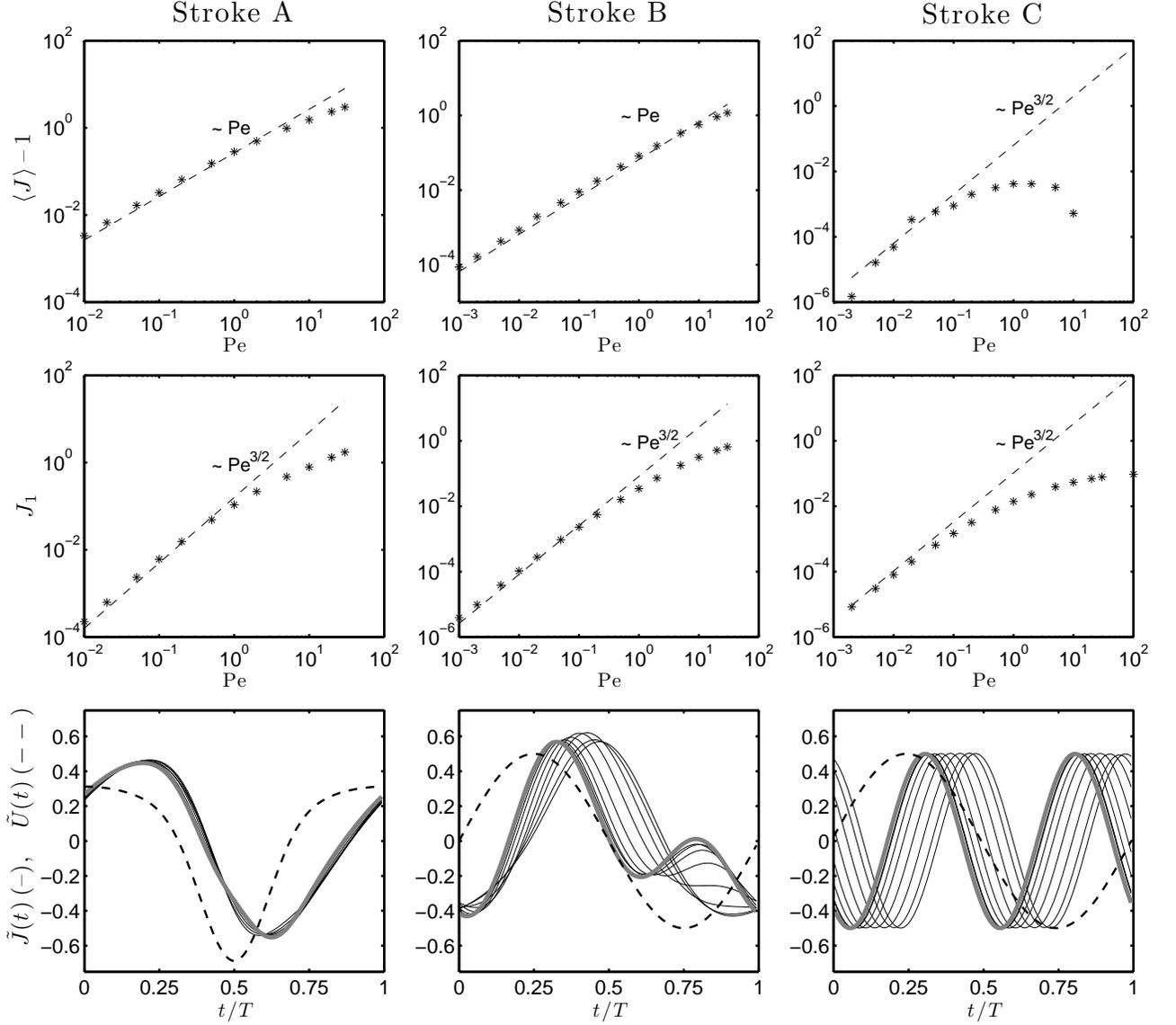}
\caption{Top: Dependence of the mean feeding rate, $\langle J\rangle$ (stars),  with $\Pe$ and comparison with the asymptotic prediction in~\eqref{eq:predict_mean} (dashed). Center:  Dependence of the peak-to-peak amplitude of the feeding rate fluctuations, $J'(t)=J(t)-\langle J\rangle$ (stars),  with $\Pe$ and comparison with the asymptotic prediction in~\eqref{eq:predict_fluct}--\eqref{eq:predict_fluct2}. Bottom: Rescaled (unit amplitude) feeding rate (solid) and velocity (dashed) time fluctuations; the asymptotic prediction for the feeding rate fluctuations at low $\Pe$ in~\eqref{eq:predict_fluct}--\eqref{eq:predict_fluct2} is shown as a thick grey line. All results are plotted for stroke A (left), stroke B (centre), and stroke C (right). }\label{fig:asympt_comp}
\end{center}
\end{figure}

It was observed previously that the value of $\Pe$ plays an important role in  the feeding ability of the cell. This is investigated further here by looking at the impact of $\Pe$ on the instantaneous feeding rate for strokes A, B and C. The instantaneous feeding rate, $J(t)$, is decomposed into its mean value, $\mean{J}$, the amplitude of its fluctuations in time, $J_1$, and its normalized profile, $\tilde{J}(t)$, so we write
\begin{equation}
J(t)=\langle J\rangle+J_1\tilde{J}(t),
\end{equation}
where $J_1=\textrm{max}(J)-\textrm{min}(J)$ and $\tilde{J}(t)=(J(t)-\langle J\rangle)/J_1$. Similar quantities are also defined for the swimming velocity: $\mean{U}$, $U_1$, and $\tilde{U}$. For a given stroke (A, B or C), the variation   of these three quantities  with $\Pe$ is displayed in  figure \ref{fig:asympt_comp}. 

For swimming strokes, it is observed that, for low $\Pe$, the modification in the mean feeding rate, $\langle J\rangle-1$, scales linearly with $\Pe$ (strokes A and B). This is consistent with the asymptotic analysis of Section~\ref{sec:asymptotics} and with the steady  results in \citet{michelin2011}. In such a diffusion-dominated regime, swimming enables the cell to sweep a region of fresher nutrients with an effective cross-section radius that is independent of the swimming velocity (because of the predominance of diffusion) and of the order of the size of the cell. At higher $\Pe$, the reduced importance of diffusion over advection reduces the effective cross-section radius and $\mean{J}$ increases at a lower rate with $\Pe$. For strokes with no net swimming motion (stroke C), the modification in the mean feeding rate scales as a higher power, $\Pe^{3/2}$, for $\Pe\leq 0.1$, consistently with the results of the asymptotic analysis. 

For both swimming and non-swimming strokes, the amplitude of the feeding rate fluctuations, $J_1$, varies as $\Pe^{3/2}$ for $\Pe\leq 1$, consistently with  our asymptotic results. On figure~\ref{fig:asympt_comp} the fluctuations profile, $\tilde{J}(t)$,  is also represented and compared to the leading order prediction of the asymptotic analysis. We see  a very good agreement at low $\Pe$ which persists even at high $\Pe$ for efficient swimming strokes such as stroke A. This confirms that the feeding rate (both its mean value and its fluctuations) is determined at leading order by the swimming mode and  corrections from the other modes only play marginal roles. Again, a clear phase delay between the swimming velocity and feeding rate is observed for all $\Pe$, and for the least efficient swimmers considered (B and C), this delay seems to increase with $\Pe$.

When $\Pe$ becomes large, another significant difference appears between strokes with zero (stroke C) or non-zero (strokes A and B) mean swimming velocity. For strokes A and B, the average feeding rate continues to increase with $\Pe$, albeit more slowly. From the  large-$\Pe$ steady  results by \citet{michelin2011}, we expect  $\mean{J}$ to scale as $\Pe^{1/2}$, when the increase in feeding rate with swimming is driven by the concentration boundary layer thickness around the cell. In contrast, for non-swimming strokes, $\mean{J}$ reaches a maximum for a finite value of $\Pe$ ($\Pe_{c1}\approx 2$) beyond which an increase in $\Pe$ actually results in a decrease of the feeding rate. Moreover, beyond a second critical value ($\Pe_{c2}\approx 11$ for this particular stroke), the mean feeding rate falls below $1$, and for large $\Pe$, swimming actually penalizes feeding as it reduces the net feeding rate below the level of the purely diffusive regime ($\Pe=0$). This somehow surprising result can be understood as follows. In stroke C, the sphere swims forward during half of a period leaving behind it a nutrient-depleted wake. In the second half of the stroke, the cell swims backward into this region of poor nutrient concentration, resulting in a reduced flux at the boundary.

\subsection{The optimal unsteady stroke is steady}
\begin{figure}
\begin{center}
\includegraphics[width=12cm]{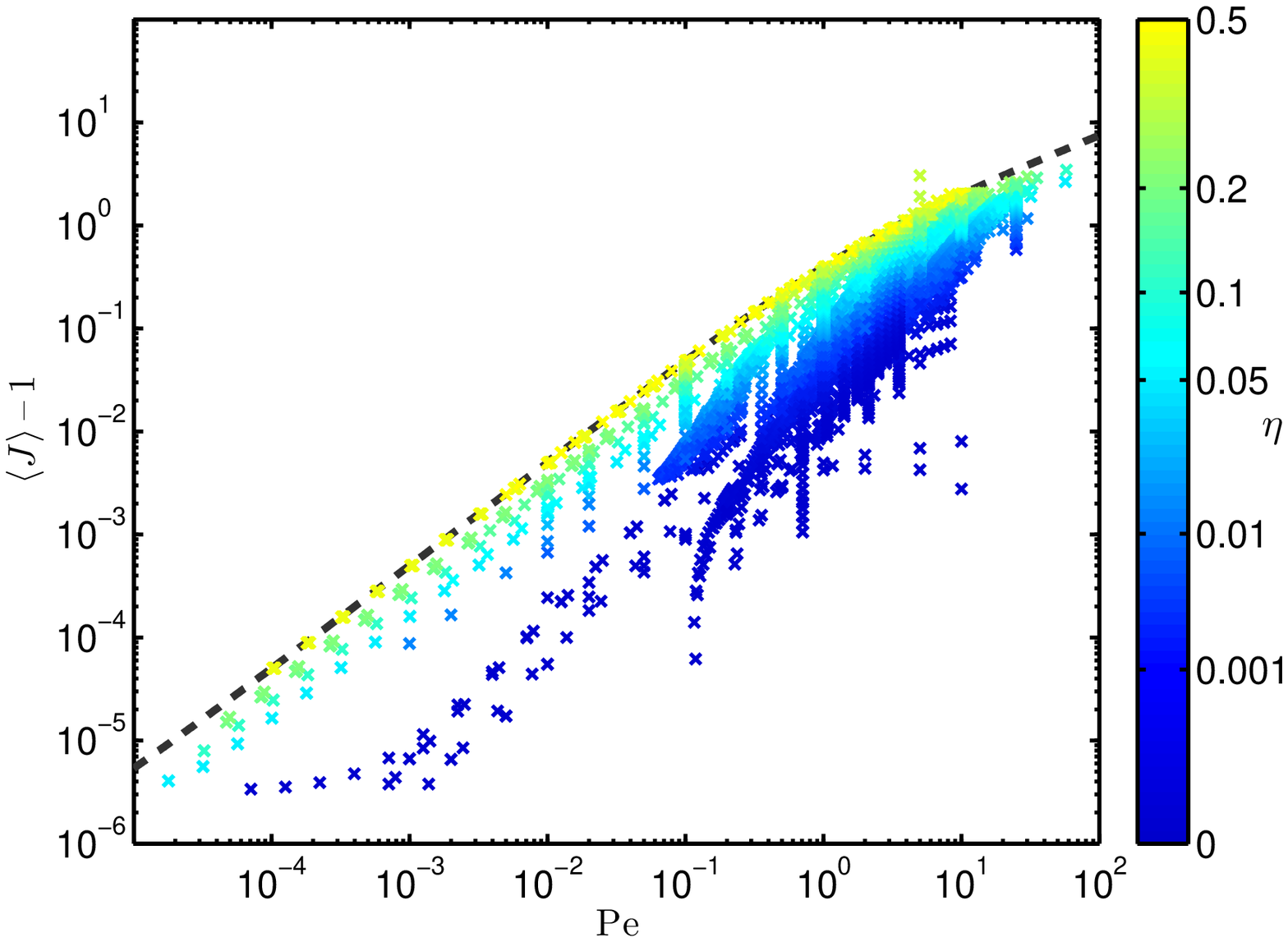}
\caption{Mean feeding rate as a function $\Pe$ for 8500 different swimming strokes (see text). The dashed line corresponds to the optimal steady feeding stroke (treadmill). For each stroke, the symbol color is related to its hydrodynamic efficiency, $\eta$.}\label{fig:J_vs_Pe}
\end{center}
\end{figure}
As we discussed above, the optimal Eulerian swimming stroke is necessarily steady. The same conclusion can not be drawn {a priori} for the feeding problem due to the time-dependence of the advection diffusion equation (see \S \ref{sec:summary}). We saw however that it was true analytically at low P\'eclet number. Numerically, it also seems to hold as illustrated in figure~\ref{fig:J_vs_Pe}. We performed numerical simulations on a large collection of unsteady Eulerian periodic and  Lagrangian periodic  strokes (8500 in total), ranging from very efficient to poor swimmers.  For all values of $\Pe$, the feeding rate is seen to be always less than that obtained with the optimal steady feeding stroke (treadmill). As for the optimal swimming stroke, the optimal Eulerian unsteady feeding stroke must therefore also be steady. Furthermore, figure \ref{fig:J_vs_Pe}  demonstrates that the more efficient the unsteady stroke is for swimming, the closer it can get to the optimal feeding rate.

\subsection{Feeding and swimming}

\begin{figure}
\begin{center}
\includegraphics[width=12cm]{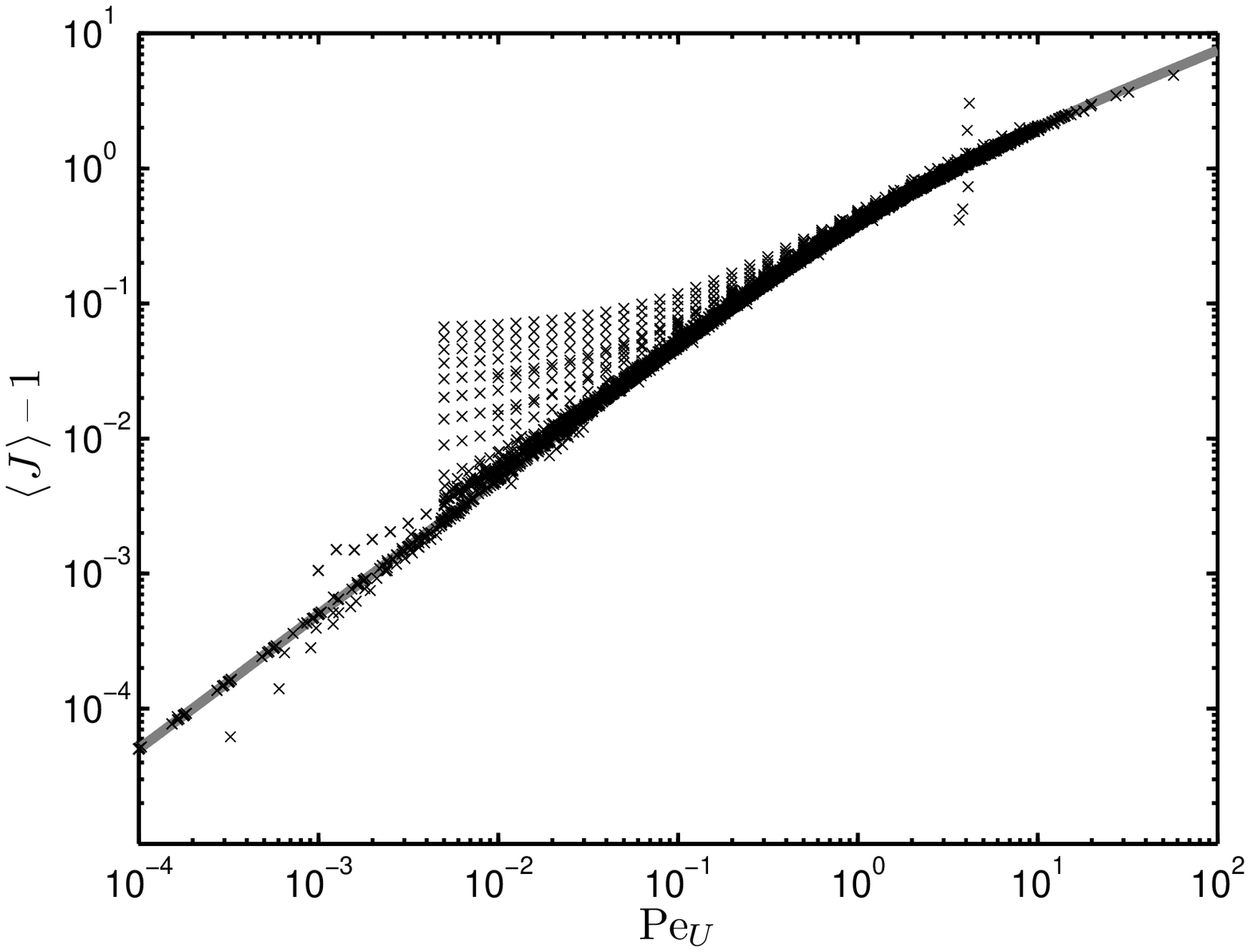}
\caption{Mean feeding rate as a function of the ``swimming P\'eclet number" $\Pe_U=\Pe\sqrt{2\eta}$ for the same 8500 strokes as in figure~\ref{fig:J_vs_Pe}. The light grey line corresponds to the feeding performance of the steady treadmill swimmer for which $\alpha_n(t)=\delta_{n,1}$.}\label{fig:J_vs_PeU}
\end{center}
\end{figure}

In the previous sections, a relationship between the swimming velocity and the feeding rate was clearly identified suggesting that at leading order, the mean feeding rate is determined by the swimming velocity and $\Pe$. More precisely, and in the light of the steady results of \citet{michelin2011}, one expects the feeding rate to be determined by the swimming P\'eclet number, $\Pe_U$, defined as
\begin{equation}
\Pe_U=\frac{a\mean{U}}{\kappa}=\Pe\sqrt{2\eta},
\end{equation}
which measures the relative importance of advection of nutrients by the net displacement of the cell and diffusion. This is clearly the case at leading order for low $\Pe$, as seen in~\eqref{eq:predict_mean}. 

In order to test the validity of this conjecture at higher $\Pe$, we plot in figure~\ref{fig:J_vs_PeU}  the mean feeding rate as a function of the ``swimming P\'eclet number" $\Pe_U$, for {the same large collection of unsteady strokes as in the previous section}. All data points collapse rather well on a single curve, that corresponds exactly to the results for the steady treadmill swimmer \citep{michelin2011}. The agreement is particularly good for larger $\Pe_U$, corresponding to more efficient swimming strokes where the swimming motion dominates. The collapse of all the data points on that curve indicates that at leading order, for all strokes and all $\Pe$, the mean feeding rate is determined by the mean swimming velocity.

Figure \ref{fig:J_vs_PeU} shows however that a significant number of points do not follow that leading order trend and are located above the grey treadmill curve. Indeed, for swimming strokes with poor efficiency (including those with  $\Pe_U=0$), the contribution from the mean swimming velocity to the mean feeding process is no longer dominant and the influence of  other squirming modes, or from time-variations of the swimming  velocity, cannot be neglected, so $\mean{J}$ remains strictly greater than one.

\section{Optimal unsteady feeding}
\label{sec:optimization}
The results presented in the previous sections and in \citet{michelin2011} suggest that (i) swimming determines feeding, at least at leading order, and as a result (ii) optimal swimming and optimal feeding strokes are essentially identical. In this section, result (ii) is confirmed directly by performing an optimization of the swimming stroke  maximizing the average nutrient uptake for a fixed energetic cost. The approach and methods presented below are based on  the frameworks presented in \citet{michelin2010c,michelin2011} and generalized here to the unsteady feeding problem for periodic Lagrangian strokes.

\subsection{Adjoint optimization framework}
 The rescaled nutrient concentration satisfies the advection-diffusion problem, \eqref{eq:advdiff}--\eqref{eq:advdiff_bc2}, and the mean feeding rate, $\mean{J}$, is given by
\begin{equation}
\mean{J}=-\Big\langle\frac{1}{4\pi}\int_\mathcal{S}\nb\cdot\grad c\,\dd S\Big\rangle,
\end{equation}
where $\nb=\eb_r$ is the outward normal unit vector. Considering a small perturbation, $\delta \ub=\sum_n\delta\alpha_n(t)\ub^{(n)}$, in the velocity field, at leading order and {for fixed $\Pe$} (or equivalently, fixed energetic cost) the resulting modification in mean feeding rate, $\mean{\delta J}=\delta\mean{J}$, is obtained at leading order as
\begin{equation}
\delta \mean{J}=-\Big\langle\frac{1}{4\pi}\int_\mathcal{S}\pard{}{n}(\delta c)\dd S\Big\rangle,
\end{equation}
where $\delta c$ is the resulting linear perturbation in the nutrient concentration field $c$ satisfying 
\begin{equation}
\varepsilon\left(\pard{}{t}(\delta c)+\ub\cdot\grad\delta c\right)-\nabla^2\delta c=-\varepsilon\delta\ub\cdot\nabla c+\frac{\delta\mathcal{P}}{2\mathcal{P}}\nabla^2 c,\label{eq:advdiff_perturb}
\end{equation}
with Dirichlet boundary conditions, $\delta c=0$, both on the surface of the swimmer and in the far-field. The last term in~\eqref{eq:advdiff_perturb} guarantees that $\Pe=\varepsilon\sqrt{\mathcal{P}}$ is constant and is obtained from $\delta\alpha_n$ and using~\eqref{eq:powerdef} as
\begin{equation}
\delta\mathcal{P}=2\sum_n\gamma_n\mean{\alpha_n\cdot\delta\alpha_n}.
\end{equation}
From~\eqref{eq:advdiff_perturb}, the change in mean feeding rate for constant $\Pe$ can be computed as 
\begin{equation}
\delta\mean{J}=\sum_n\mean{\tilde{\alpha}_n\cdot\delta\alpha_n},
\end{equation}
where
\begin{equation}
\tilde\alpha_n(t)=\alpha_n^*(t)-\frac{\gamma_n\alpha_n(t)}{\mathcal{P}}\left(\mean{J}-\mathcal{H}\right),
\end{equation}
is the gradient of the feeding rate, at constant $\Pe$, with respect to the $n$-th squirming mode amplitude and
\begin{equation}
\alpha_n^*(t)=\frac{\varepsilon}{4\pi}\int_{\Omega_f}g\grad c\cdot\ub^{(n)}\dd \Omega,\quad
\mathcal{H}=\Big\langle\frac{1}{4\pi}\int_{\Omega_f}\grad c\cdot\grad g\,\dd\Omega\Big\rangle.
\end{equation}
In the previous equation, $\Omega_f$ is the entire fluid domain, $\ub^{(n)}$ is the steady velocity field of the $n$-th squirming mode, and the adjoint field, $g$, satisfies the adjoint advection-diffusion problem
\begin{equation}\label{eq:adjoint}
\varepsilon\left(\pard{g}{t}+\ub\cdot\grad g\right)=-\nabla^2 g,
\end{equation}
together with boundary conditions
\begin{align}
g\rightarrow 0, &\textrm{   for   } r\rightarrow \infty,\\
g=1,&\textrm{   for   } r=1.
\end{align}

A given Lagrangian periodic swimming stroke is defined by the trajectories of the surface material points, $\xi(\mu_0,t)$. The gradient of $\mean{J}$ with respect to the stroke, $\xi(\mu_0,t)$, is then the unique function $F[\xi](\mu_0,t)$ such that for any stroke perturbation $\delta\xi$, the resulting modification in $\mean{J}$ is 
\begin{equation}
\delta\mean{J}=\frac{1}{2\pi}\int_0^{2\pi}\int_{\Omega_f}F[\xi](\mu_0,t)\, \delta\xi(\mu_0,t)\,\dd\mu_0\dd t.
\end{equation}
This gradient can be obtained directly from $\tilde{\alpha}_n(t)$ as
\begin{equation}
F[\xi](\mu_0,t)=\frac{1}{2}\left[\tilde\alpha_nL'_n(\xi)\pardd{\xi}{\mu_0}{t}+\pard{}{t}\left(\tilde\alpha_nL_n'\pard{\xi}{\mu_0}\right)\right],
\end{equation}
and then projected onto the subspace of acceptable strokes (periodic trajectories, no displacement at the pole) \citep{michelin2010c}. Note that although presented here for the particular case of a spherical swimmer, this optimization framework can easily be generalized to periodic swimming strokes of organisms with arbitrary shapes \citep{michelin2011}.

\subsection{Optimal feeding strokes}

Following \citet{michelin2010c} and in order to account for constraints on the stroke kinematics (introduced for example by a finite cilia-length-to-cell-size ratio), an additional constraint is included in the optimization algorithm to limit the maximum amplitude of angular displacements, $\Theta_\textrm{max}$, of any surface point during the stroke. This optimization is performed using a steepest-ascent iterative optimization algorithm as described by \citet{michelin2011}, and the gradient of the feeding rate with respect to the swimming stroke is computed using the  results from previous sections.

\begin{figure}
\begin{center}
\includegraphics[width=.95\textwidth]{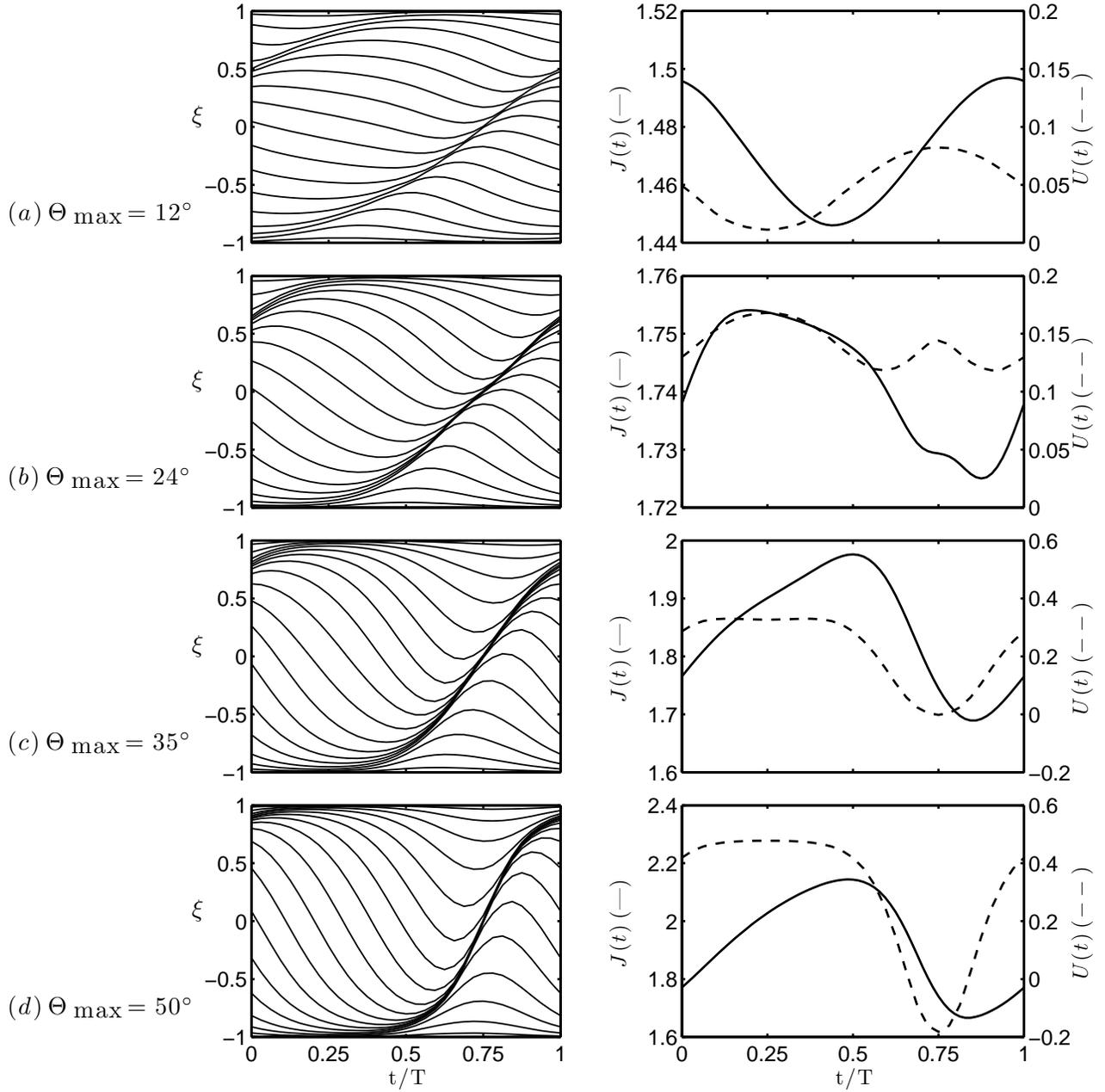}
\caption{Left: Lagrangian trajectories of four optimal feeding strokes, $\xi(t)$, obtained for $\Pe=5$ and a maximum angular stretching of the surface equal to (a) $\Theta_\textrm{max}=12^\circ$, (b) $\Theta_\textrm{max}=24^\circ$, (c) $\Theta_\textrm{max}=35^\circ$, and (d) $\Theta_\textrm{max}=50^\circ$. Right: Time-variation for each of these optimal strokes of the instantaneous feeding rate, $J(t)$ (solid), and swimming velocity, $U(t)$ (dashed). The characteristics of these four strokes are summarized in   Table~\ref{tab:optimPeP5}.
}\label{fig:optim_traj_Pe5}
\end{center}
\end{figure}

\begin{figure}
\begin{center}
\begin{tabular}{cc}
\includegraphics[width=.45\textwidth]{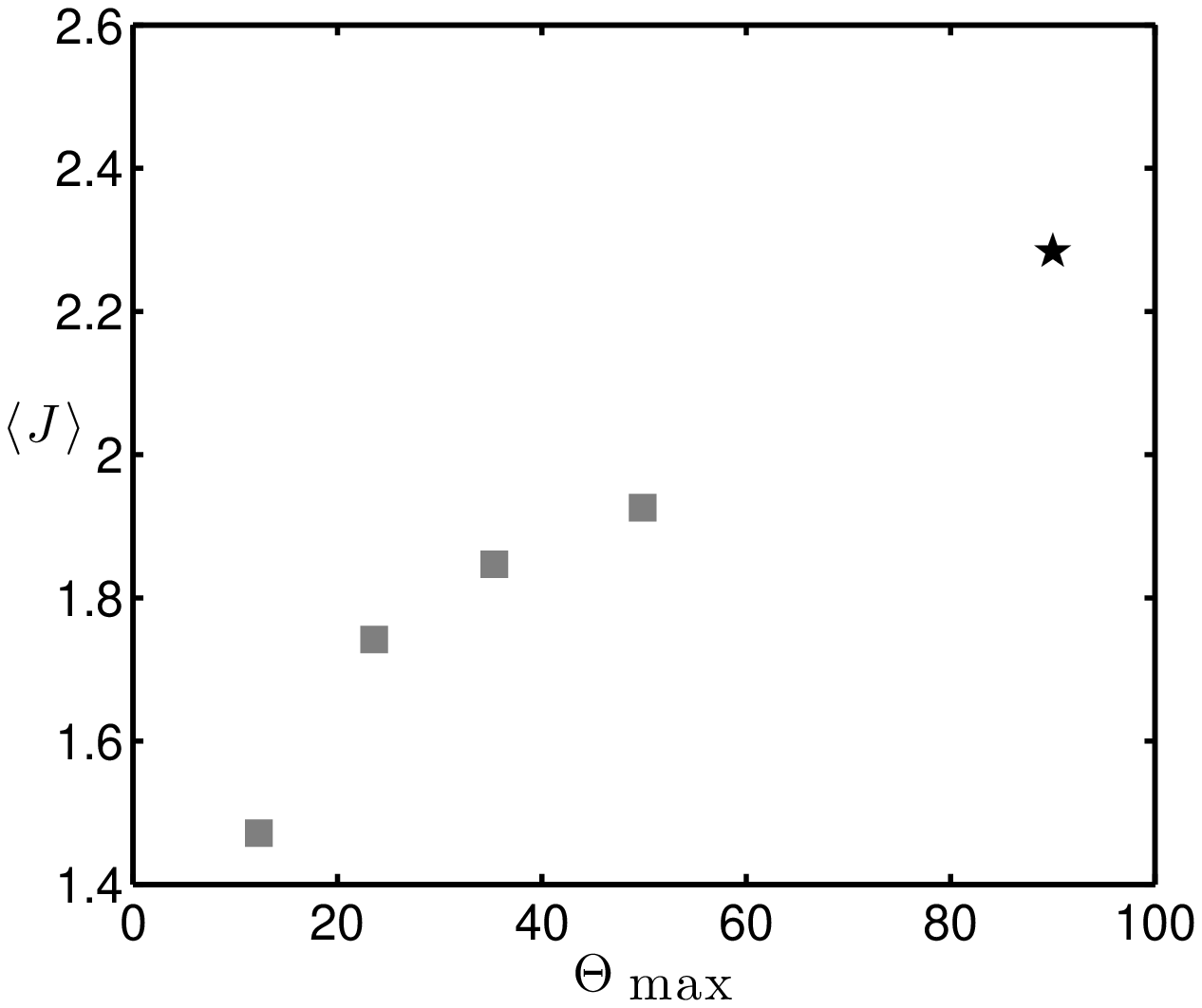} &
\includegraphics[width=.45\textwidth]{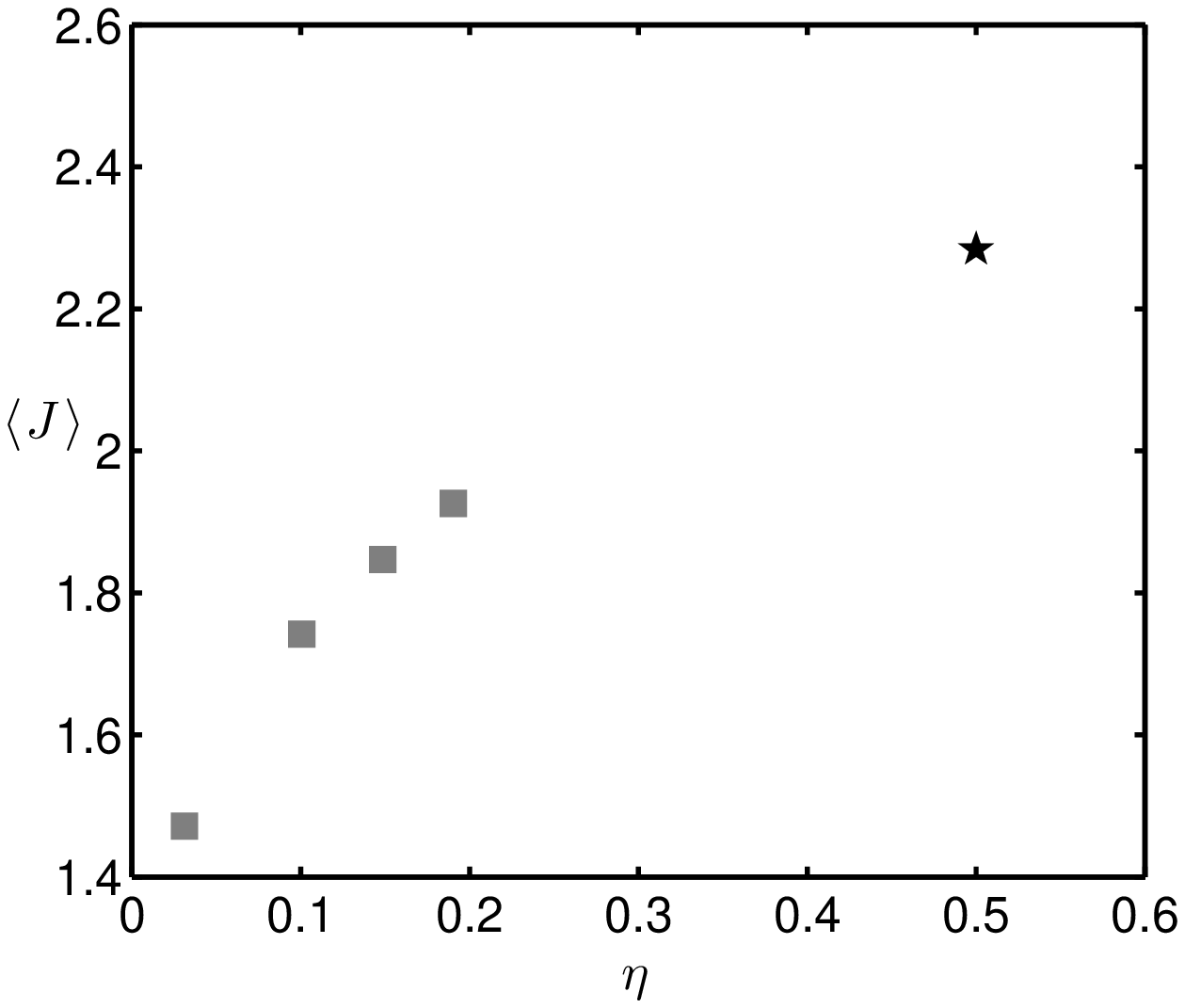}
\end{tabular}
\caption{Left: Maximum feeding rate, $\langle J \rangle$,  as a function of the maximum angular displacement angle, $\Theta_\textrm{max}$,  in optimal strokes for $\Pe=5$. Right: Maximum feeding rate as a function of swimming efficiency, $\eta$, for optimal feeding strokes obtained for $\Pe=5$ and various maximum displacement angles $\Theta_\textrm{max}$. The black stars in both figures correspond to the optimal steady stroke (treadmill).}\label{fig:optimPeP5}
\end{center}
\end{figure}

Figure \ref{fig:optim_traj_Pe5} shows the optimal strokes obtained for $\Pe=5$ and four increasing values of  $\Theta_\textrm{max}$. The optimal strokes consist in two different parts: an effective stroke where the surface of the squirmer stretches from front to back, enabling the swimming motion, followed by a recovery stroke where the material points (e.g. cilia tips) accumulated in the back side of the sphere are brought back to their original position with a front-like dynamics, reminiscent of the metachronal waves observed in ciliates. A wave velocity can be defined from the synchronization of the trajectories \citep{michelin2010c}. Notice also in figure \ref{fig:optim_traj_Pe5} the phase delay between feeding and swimming predicted theoretically.  Imposing tighter bounds on $\Theta_\textrm{max}$ results in a slower phase-velocity of the recovery stroke, in a smaller and steadier swimming velocity, and in a reduced efficiency (Table \ref{tab:optimPeP5}). This dichotomy of the optimal stroke and impact of the maximum displacement $\Theta_\textrm{max}$ are essentially identical to that observed in the optimal Lagrangian swimming stroke by \citet{michelin2010c}, for which it was observed that a continuous set of optimal strokes could be obtained for $0\leq\Theta_\textrm{max}\leq 90^\circ$, approaching asymptotically the optimal steady  swimmer when $\Theta_\textrm{max}\rightarrow 90^\circ$. A similar behavior is observed on Figure~\ref{fig:optimPeP5}(a).

\begin{table}
\begin{center}
\begin{tabular}{cccccccc}
&$\quad\Theta_\textrm{max}\,(^\circ)\quad$ & $\quad\mean{J}\quad$ & $\quad\mean{U}\quad$ & $\quad\eta\quad$ & $\quad\Pe\quad$ & $\quad\varepsilon\quad$ & $\quad\Pe_U\quad$\\
\\
(a) &$12$ & $1.47$ & $0.048$ & $3.1\%$ & $5$ &$26.2$ & $1.25$\\
(b) &$24$ & $1.74$ & $0.141$ & $10\%$ & $5$ &$15.9$ & $2.24$ \\
(c) &$35$ & $1.85$ & $0.221$  & $15\%$ &$5$ & $12.3$ & $2.73$ \\
(d) &$50$ & $1.93$ & $0.295$ & $19\%$ &$5$ & $10.5$ & $3.09$\\
\\
\end{tabular}
\caption{Characteristics of the optimal feeding strokes obtained computationally for $\Pe=5$ and four maximum angular displacements, $\Theta_\textrm{max}$, displayed on figure~\ref{fig:optim_traj_Pe5}.}\label{tab:optimPeP5}
\end{center}
\end{table}

\begin{figure}
\begin{center}
\includegraphics[width=.95\textwidth]{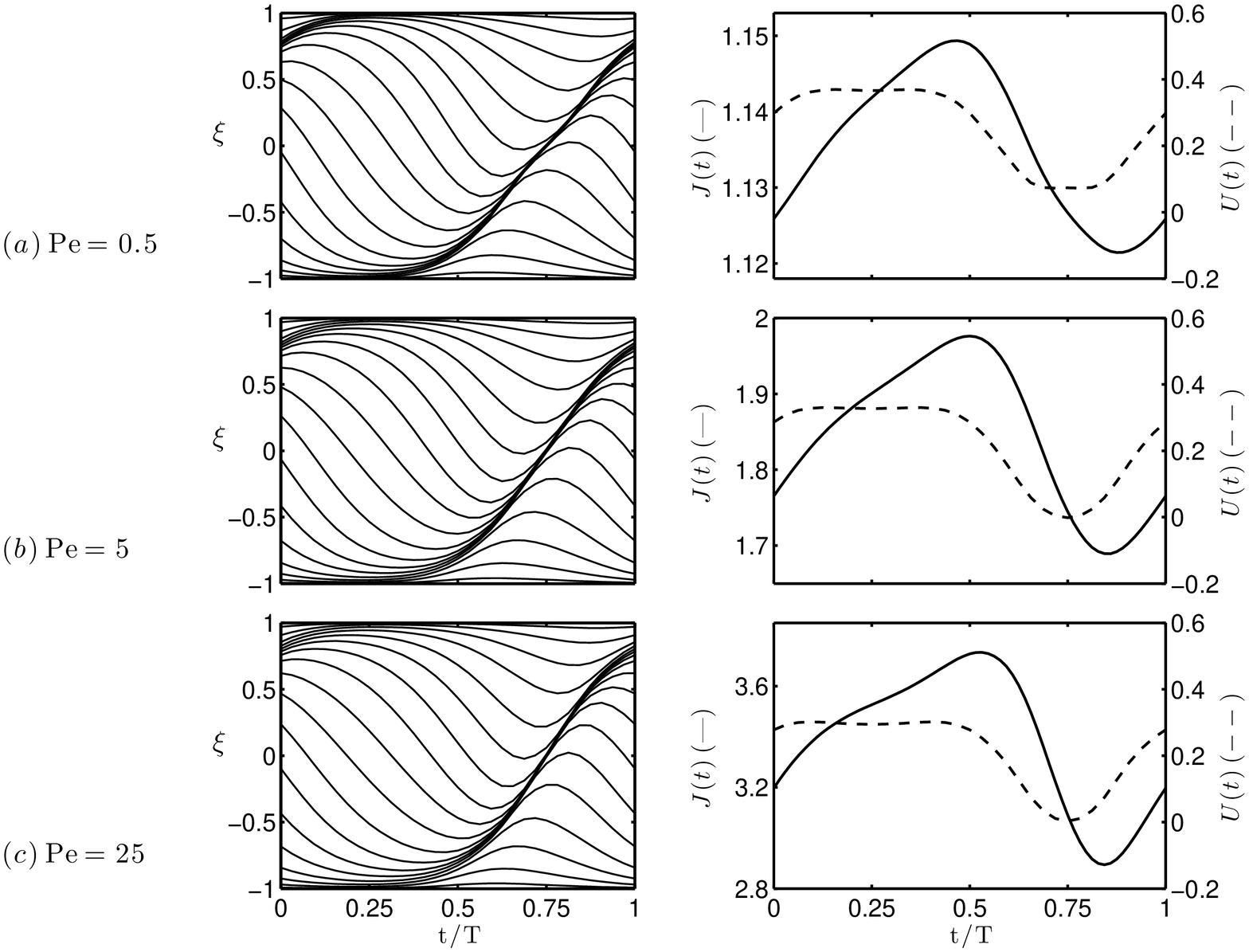}
\caption{Left: Lagrangian trajectories of the optimal  time-periodic feeding strokes obtained for $\Theta_\textrm{max}=35^\circ$ and  (a) $\Pe=0.5$, (b) $\Pe=5$ and (c) $\Pe=25$. Right: Time-variation for each of these optimal strokes of the instantaneous feeding rate, $J(t)$ (solid), and swimming velocity, $U(t)$ (dashed). 
}\label{fig:optim_traj_TH35}
\end{center}
\end{figure}

The above conclusions are unchanged when performing the optimization at different values of the P\'eclet number, as shown in figure \ref{fig:optim_traj_TH35}.  For a given constraint on the maximum displacement $\Theta_\textrm{max}$, the same strokes are obtained regardless of the value of $\Pe$. These results confirm therefore that the optimal unsteady feeding stroke is essentially the same as the optimal swimming stroke, regardless of the value of the P\'eclet number. In both cases (swimming or feeding), the optimal Lagrangian stroke can be understood as a periodic approximation of the optimal steady stroke.

\section{Conclusions}
\label{sec:conclusions}

In this paper we use asymptotic analysis and numerical computations to address the link between swimming and feeding for motile microorganisms. Using the mathematical model of spherical squirmers acting on the viscous fluid in a time-periodic manner, we first show analytically at low $\Pe$ that the mean rate of feeding as well as its fluctuations in time depend only on the  swimming modes of the squirmer up to order $\Pe^{3/2}$, even when no swimming occurs on average, while the influence of  non-swimming modes come  in later at order $\Pe^2$. We also demonstrate the existence of a phase delay between feeding and swimming of 1/8th of a period.  Using three illustrative stokes, we then employ numerical computations to confirm our asymptotic results and further demonstrate the relationship between  swimming and feeding. Using 
 adjoint-based optimization we finally determine numerically the optimal unsteady strokes maximizing feeding rate for a fixed energy budget.  The overall optimal is always the steady swimmer. For time-periodic strokes, we find - as in the steady case - that the optimal  feeding strokes are equivalent to those optimizing swimming - this result is true for  all P\'eclet numbers even though the value of feeding rate strongly depends on the P\'eclet number. As for the optimal unsteady swimming problem, optimal feeding strokes are therefore mathematical regularizations of the steady  problem (treadmill) of overall maximum swimming and feeding performance.

Clearly the problem studied here is idealized in many ways. The geometry is that of a sphere and the boundary conditions assume perfect nutrient absorption. These simplifications allow us however to develop a precise mathematical and computational description of the  problem, both for the fluid and for the passive nutrient concentration. It is hoped that the biophysical insight developed in this study will be applicable to a wide range of problems in the realm of microorganism locomotion,  e.g.~in bacterial chemotaxis (at low $\Pe$) or the feeding of plankton (at high $\Pe$). One of the main modeling challenge for future work concerns the issue of shape changes. Most motile organisms display a Lagrangian deformation of their shapes. In this paper we have assumed that the deformations (the surface boundary conditions) always act tangentially to the organism surface, allowing the shape to remain that of a sphere. Clearly normal surface velocities would also need to be considered, and these are precisely the ones leading to changes in shape. The problem would then involve solving for the flow and nutrient concentration around a time-varying boundary. We hope that our study will inspire future work in this direction.

\section*{Acknowledgments}
This work was supported in part by the US National Science Foundation through grant CBET-0746285.

\appendix
\section{Definition of the $A_{mnp}$ and $B_{mnp}$ tensors}
\label{sec:AB}
The coefficients $A_{mnp}$ and $B_{mnp}$ used in~\eqref{eq:advdiff_discrete} are defined in terms of the Legendre polynomials as follow:
\begin{align}
A_{mnp}=&\frac{(2p+1)(2n+1)}{2}\int_{-1}^1L_m\,L_n\,L_p\,\dd\mu,\label{eq:defA}\\
B_{mnp}=&\frac{(2p+1)(2n+1)}{2n(n+1)}\int_{-1}^1(1-\mu^2)L'_m\,L'_n\,L_p\,\dd \mu.\label{eq:defB}
\end{align}
They are easily computed using 
\begin{equation}
A_{m0p}=\delta_{mp},\qquad B_{m0p}=0.
\end{equation}
and the following recursive relations for $n\geq 1$
\begin{align}
A_{mnp}=&\frac{2n+1}{n}\left[-\frac{n-1}{2n-3}A_{m,n-2,p}+\frac{m+1}{2m+1}A_{m+1,n-1,p}+\frac{m}{2m+1}A_{m-1,n-1,p}\right],\\
B_{mnp}=&\frac{2n+1}{n(n+1)}\left[\frac{(n-2)(n-1)}{2n-3}B_{m,n-2,p}+\frac{m(m+1)}{2m+1}\left(A_{m-1,n-1,p}-A_{m+1,n-1,p}\right)\right].
\end{align}

\section{Unsteady heat/mass transfer around a sphere in Stokes flow}
 \label{sec:acrivos_unsteady}
 In Section \ref{sec:asymptotics}, the asymptotic expansion of the concentration distribution around a general squirmer and the resulting feeding rate, $J(t)$, were obtained in the limit $\Pe\ll 1$. The results obtained in \eqref{eq:predict_fluct}--\eqref{eq:predict_fluct2} also hold for any spherical object moving at velocity $U(t)$, regardless of whether the sphere is swimming (zero net force) or a rigid sphere actuated by an external force, as we now show.
 
 Indeed, considering a generalization of the work of \citet{acrivos1962} to unsteady particle velocity $\alpha_1(t)=U(t)$, the velocity field around the sphere is given by the streamfunction
 \begin{equation}
 \label{eq:rigidsphere}
 \psi(r,\mu,t)=\frac{\alpha_1(t)(1-\mu^2)}{2}\left(\frac{3r}{2}-r^2-\frac{1}{2r}\right).
 \end{equation}
 Following the same approach as in \S\ref{sec:asymptotics},~\eqref{eq:advdiff_nearfield} takes the same form but~\eqref{eq:advl1}--\eqref{eq:advln} become
 \begin{align}
 l_1&=-\left(1-\frac{3}{2r}+\frac{1}{2r^3}\right)\mu\pard{}{r}-\frac{1-\mu^2}{r}\left(1-\frac{3}{4r}-\frac{1}{4r^3}\right),\\
 l_n&=0\quad \textrm{for all    }n\geq 2.
 \end{align}
 In the same way,~\eqref{eq:advdiff_UBL}--\eqref{eq:advdiff_SBL} are slightly modified due to the contribution of the Stokeslet in the far-field:
\begin{align}
&\textrm{in the UBL,}\qquad \mathcal{D}\cdot \mathcal{C}_p=\ci p\mathcal{C}_p+\varepsilon^{1/2}\sum_{q=-\infty}^\infty\alpha_{1,q}\mathcal{L}_1\cdot \mathcal{C}_{p-q} \nonumber\\
&\hspace{5cm}+\varepsilon\sum_{q=-\infty}^\infty\alpha_{1,q}\tilde{\mathcal{L}}_1\cdot\mathcal{C}_{p-q}+O(\varepsilon^{3/2}),\label{eq:advdiff_UBL2}\\
\nonumber\\
&\textrm{in the SBL,}\qquad\mathscr{D}\cdot \mathscr{C}_0=\alpha_{1,0}\,\mathscr{L}_1\cdot \mathscr{C}_0 +\varepsilon\alpha_{1,0}\,\tilde{\mathscr{L}}_1\cdot \mathscr{C}_0 +O(\varepsilon^2)\label{eq:advdiff_SBL2},
\end{align}
where $\mathcal{L}_1$ and $\mathscr{L}_1$ remain unchanged from~\eqref{eq:adv_op_bl}, and 
\begin{equation}
\tilde{\mathcal{L}}_1=\frac{3\mu}{2R}\pard{}{R}+\frac{3(1-\mu^2)}{4R^2}\pard{}{\mu},
\end{equation}
 and $\tilde{\mathscr{L}}_1$ is obtained by replacing $R$ by $\rho$ in the previous equation.
 
 Following the approach of \S\ref{sec:asymptotics}, equations~\eqref{eq:bc_surface}--\eqref{eq:predict_fluct2} remain unchanged except:
 \begin{itemize}
 \item{Equation~\eqref{eq:modif1} becomes
 \begin{equation}
 D\cdot c_p^2=\alpha_{1,p}\,l_1\cdot c_0^0=\frac{\mu\alpha_{1,p}}{r^2}\left(1-\frac{3}{2r}+\frac{1}{2r^3}\right), 
 \end{equation}
 }
 \item{Equation~\eqref{eq:modif2} becomes
 \begin{equation}
c_p^2(r,\mu)=\alpha_{1,p}\mu\left(-\frac{1}{2}+\frac{3}{4r}+\frac{1}{8r^3}-\frac{3}{8r^2}\right)+\sum_{n=1}^\infty \gamma_{n,p}L_n(\mu)\left(\frac{1}{r^{n+1}}-r^n\right),
\end{equation}
 }
 \item{Equation~\eqref{eq:res_ordre2} becomes
 \begin{equation}
 c_p^2=\frac{\alpha_{1,0}}{2}\left(\frac{1}{r}-1\right)\delta_{p,0}+\mu\alpha_{1,p}\left(-\frac{1}{2}+\frac{3}{4r}+\frac{1}{8r^3}-\frac{3}{8r^2}\right),
 \end{equation}
 }
 \item{and Equation~\eqref{eq:modif4} becomes
 \begin{align}
\frac{1}{R^2}\totd{}{R}\left(R^2\totd{\tilde{\mathcal{C}}_p^3}{R}\right)-\ci p \tilde{\mathcal{C}}_p^3&=\frac{1}{2}\sum_{q=-\infty}^\infty\alpha_{1,p-q}\int_{-1}^1\mathcal{L}_1\cdot\mathcal{C}_p^2\dd\mu+\frac{1}{2}\sum_{q=-\infty}^\infty\alpha_{1,p-q}\int_{-1}^1\tilde{\mathcal{L}}_1\cdot\mathcal{C}_p^1\dd\mu\nonumber\\
&=\frac{1}{3R}\sum_q\alpha_{1,p-q}\alpha_{1,q}\ee^{-R\sqrt{\ci q}}.
\end{align}
}
 \end{itemize}
 
These modifications do not impact the final result for the nutrient flux at the boundary. The expansion of the feeding rate for an oscillating sphere is therefore identical to that of the squirmer with same swimming velocity up to $O(\Pe^{3/2})$ in~\eqref{eq:predict_fluct}--\eqref{eq:predict_fluct2}. Looking at the corrections in the asymptotic expansion presented above, it appears that any far-field singularity in the velocity field (Stokeslet, etc...) will modify the near-field solution starting at $O(\Pe)$ and the unsteady boundary layer $O(\Pe^{3/2})$ but that such modifications will only affect the azimuthal fluctuations of the concentration and not its azimuthal average which determines the total feeding rate. Therefore, the asymptotic expansion of the feeding rate remains unchanged for any sphere moving at velocity $\alpha_1(t)$, regardless of the tangential velocity field applied on its surface, and regardless of the total force applied on the sphere.
 
 As a result, equations~\eqref{eq:predict_fluct}--\eqref{eq:predict_fluct2} are a generalization to unsteady motions of the classical result on the heat and mass transfer on a sedimenting sphere \citep{acrivos1962}, and the physical conclusions of \S\ref{sec:asymptotics} are also valid in the case of a rigid sphere, in particular (i) the phase delay between the velocity and the mass transfer rate and (ii) an  increase in mass/heat transfer scaling as $\Pe^{3/2}$  for a sphere oscillating around a fixed mean position ($\langle U \rangle =0$).


\end{document}